\documentclass[twocolumn,showpacs,aps,groupaddress]{revtex4}
\usepackage[]{graphicx}
\usepackage{times}
\begin{document}
\preprint{a manuscript for a review article,}
\title{Optical Properties of Excitons in ZnO-based Quantum Well Heterostructures}

\author{T. Makino}
\email[electronic mail: ]{tmakino@postman.riken.go.jp}
\affiliation{Photodynamics Research Center, RIKEN (The Institute of Physical and Chemical Research), Aramaki-aza-Aoba 519-1399, Sendai 980-0845, Japan}
\author{Y. Segawa}
\altaffiliation{Department of Physics, Tohoku University, Sendai 980-8577, Japan}
\affiliation{Photodynamics Research Center, RIKEN (The Institute of Physical and Chemical Research), Aramaki-aza-Aoba 519-1399, Sendai 980-0845, Japan}
\author{M. Kawasaki}
\altaffiliation{Also a member of Combinatorial Material Exploration, Japan Science and Technology Corporation, Tsukuba, Japan}
\affiliation{Institute for Materials Research, Tohoku University,
Sendai 980-8577, Japan}
\author{H.~Koinuma}
\altaffiliation{Also a member of Combinatorial Material Exploration, Japan Science and Technology Corporation, Tsukuba, Japan}
\affiliation{Materials and Structures Laboratory, Tokyo Institute of Technology, Yokohama 226-8503, Japan}

\date{\today}

\begin{abstract}
Recently the developments in the field of II-VI-oxides have been spectacular. Various epitaxial methods has been used to grow epitaxial ZnO layers. Not only epilayers but also sufficiently good-quality multiple quantum wells (MQWs) have also been grown by laser molecular-beam epitaxy (laser-MBE). We discuss mainly the experimental aspect of the optical properties of excitons in ZnO-based MQW heterostructures. Systematic temperature-dependent studies of optical absorption and photoluminescence in these MQWs were used to evaluate the well-width dependence and the composition dependence of the major excitonic properties. Based on these data, the localization of excitons, the influence of exciton-phonon interaction, and quantum-confined Stark effects are discussed.

The optical spectra of dense excitonic systems are shown to be determined mainly by the interaction process between excitons and biexcitons. The high-density excitonic effects play a role for the observation of room-temperature stimulated emission in the ZnO MQWs. The binding energies of exciton and biexciton are enhanced from the bulk values, as a result of quantum-confinement effects.
\end{abstract}
\maketitle

\section{Introduction}

ZnO has a large fundamental band gap of $\sim $3.37~eV at room
temperature~\cite{makino8,chen1,gil2}. The nature of high thermal
conductivity, high luminous efficiency and mechanical and chemical
robustness have made ZnO and its alloys the promising material
system for light-emitting devices operated at a UV spectral region~\cite{segawa1}. New applications for ZnO field effect transistors
are also under extensive exploration~\cite{nomura1,carcia1,nishii1,masuda1,hoffman1}. In addition to these
stabilities, ZnO has the advantage of a larger exciton binding
energy~\cite{LBZincoxide} (about 60~meV), which assures more efficient excitonic
emission at higher temperatures. Moreover, the excitons in ZnO-based
quantum well (QW) heterostructures exhibit strong stability as
compared to bulk semiconductors or III-V QWs due to the enhancement
of the binding energy~\cite{sun1,sunfull1} and the reduction of the exciton-phonon
coupling~\cite{sun2} caused by quantum confinement. Due to these effects,
excitons are expected to play an important role in many-body
processes such as laser action and nonlinear absorption of II-VI-oxide QWs even at room temperature. For example, excitonic gain has
been demonstrated in ZnO/ZnMgO QWs. A detailed study of excitons in
ZnO multiple quantum wells (MQWs) is thus important to understand
the optical properties of these wide gap heterostructures, also in
view of their application to ultraviolet-blue optoelectronic devices~\cite{makino11}. In addition,
an effect of built-in electric fields inside QW layers might be taken into account for
ZnO QWs having relatively high barrier height. As has been extensively investigated in wurtzite GaN-related heterostructures, in biaxially
strained wurtzite heterolayers with
the \textit{c} axis parallel to the growth direction, piezoelectric and spontaneous polarization effects may be present as a
consequence of its noncentrosymmetry~\cite{bernardini1,seoung-hwanpark,cingolani2,langer1,m_leroux1}.

In this review article, we overview
the most recent experimental and theoretical work about the optical
properties of excitons in ZnO-based QWs grown on lattice-matched
ScAlMgO$_4$ (SCAM) substrates. We briefly introduce various wide-gap
materials that have been used to construct the ZnO-related double
heterostructures in section~\ref{sec:combination}. In section \ref{sec:excitontheory} we
briefly summarize the basic theoretical concepts used to model quasi
two-dimensional excitons in QWs. In sections \ref{sec:QCSE} and \ref{sec:linear} we
discuss the linear optical properties of excitons, including the
thermal stability, the strength of the quantum-size and the quantum-confined Stark effects in QWs,
and the temporal evolution of the excitonic transitions. In section
\ref{sec:nonlinear} we treat the nonlinear optical properties of excitons.
After a short theoretical introduction, we discuss the role of excitons in the stimulated emission
processes of II-VI-oxide QWs in subsection~\ref{sec:biexciton}.
Finally, we also discuss the enhancement in biexciton binding energy and the stimulated emission from two-component carrier plasmas (i.e., electron-hole plasmas). Our conclusions
are drawn in section \ref{sec:conclusion}.

\section{Various zinc oxide-related double heterostructures}
\label{sec:combination}
Various wide-gap semiconducting or insulating materials have been adopted to construct ZnO-related heterostructures. Since this article is devoted to the optical properties of the QWs, we concentrate on the materials adopted for the double heterostructures. Amongst them, the ZnO/ZnMgO double heterostructure is a quantum structure that has been most extensively studied with spectroscopic characterization. Resonant tunneling action was observed in a similar QW~\cite{krishnamoorthy}. Because the stable crystal symmetry of ZnO is different from that
of MgO or CdO, the growth of these ternary alloys over the whole concentration range is rather difficult. We should here emphasize the almost perfect lattice mismatch attained between CdZnO and MgZnO. This ternary-ternary QW can realize a perfect (in-plane) lattice-match with a maximum
barrier height up to 0.9~eV by choosing an appropriate
combination of cadmium and magnesium concentrations (cf. Fig.~4 in Ref.~\cite{makino14})~\cite{kawasaki2,makino14}. This is an advantage compared to (In,Ga)N/(Al,Ga)N
QWs. Not only wurtzite nitrides~\cite{vispute1,narayan1} such as GaN or AlN but also oxides-based insulators are found to be also good candidates for the barriers of the QWs. Ohkubo \textit{et~al.}~\cite{ohkubo1} reported the growth and the the optical characterization of $\beta $-LiGaO$_2$/ZnO/ScAlMgO$_4$. Fujimura \textit{et~al.}~\cite{edahiro1} reported growth of the heterostructures based on diluted magnetic semiconductors, ZnO/ZnMnO. Bogatu \textit{et~al.} claimed the existence of surface QW-like states in hydrogen-implanted ZnO crystals~\cite{bogatu1}.

\section{Modeling of excitonic states in {ZnO} quantum wells}
\label{sec:excitontheory}
\begin{figure}
	\includegraphics[width=.4\textwidth]{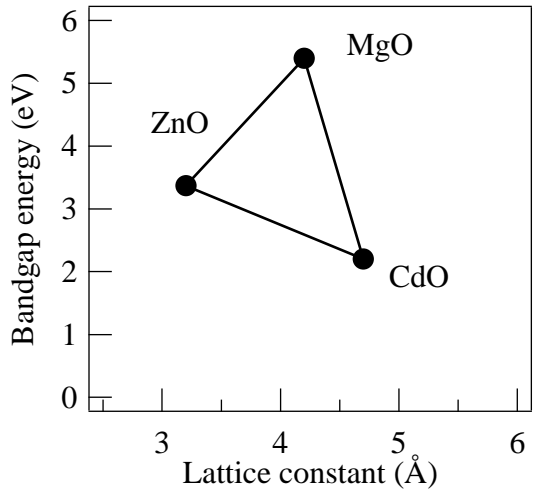}
	\caption{Energy gap at 300 K versus lattice constant for a few relevant II-VI-oxide semiconductors.}
	\label{gapvslattice}
\end{figure}
In this review article, we concentrate on II-VI oxide QWs consisting of binary compounds ZnO, MgO, CdO and their solid solutions ZnMgO and ZnCdO because many of spectroscopic studies have been conducted in these systems~\cite{chen-jvb}. The combination of these isoelectronic materials (ZnO, MgO, CdO) spans the whole blue ultraviolet spectral range (Fig.~\ref{gapvslattice}) through with the complication of lattice mismatch which strongly affects the electronic states and the valence band offset.
\begin{table}[htb]
	
	\caption{Main structural and electronic parameters of binary II-VI-oxide compounds}
	\begin{center}
		\begin{ruledtabular}
			\begin{tabular*}{\hsize}{l@{\extracolsep{0ptplus1fil}}c@{\extracolsep{0ptplus1fil}}c@{\extracolsep{0ptplus1fil}}r}
				 &ZnO\tablenotemark[1]&MgO&CdO\\
				\hline \\
				$E_g$ (eV)&3.37&5.4&2.2\\
				$m_e (m_0)$&0.28&--&--\\
				$m_{hh} (m_0)$&0.78&--&--\\
				$a$ (\textrm{\AA})&3.2&4.2&4.7\\
			\end{tabular*}
	\tablenotetext[1]{from Ref.~\onlinecite{LBZincoxide}.}
		\end{ruledtabular}
	\end{center}
		\label{tbl:physicalparam}
\end{table}

The calculation of the confinement energies and excitonic states requires a detailed knowledge of the band structure parameters. We should mention that most of the band parameters of ZnO
and MgO and related ternary alloy are not well known. Those used in this article~\cite{LBZincoxide} are summarized in Table~\ref{tbl:physicalparam}. It should be pointed out that Mg$_x$Zn$_{1-x}$O alloys grow with wurtzite symmetry and physical parameters available in literature are for MgO which grows with zincblende symmetry thus making it impossible to determine the values of physical parameters in the alloy system by interpolation. Because the effective masses or dielectric constants are not known for MgO, those for the solid solutions are not either known. On the other hand, the band gap energy in Mg$_{x}$Zn$_{1-x}$O have been given
elsewhere~\cite{matsumoto1}. Coli and Bajaj~\cite{coli1} deduced a reliable band gap offset since there have been so far no
reliable experimental determinations of the ratio between the
conduction and valence band offsets ($\Delta E_c / \Delta E_v$) in these
heterostructures~\cite{ohtomo4,makino11}. They have used $\Delta E_c / \Delta E_v$ as a fitting parameter,
varying it from 90/10 to 60/40 and found
that their results are closer to the experimental data when the
ratio is in the range 60/40--70/30. The electronic states are evaluated in the effective mass approximation.

The exciton binding energy is a very important parameter to be evaluated in II-VI-oxide compounds. The enhancement of the excitonic stability due to quantum confinement~\cite{sun1,sunfull1} and reduced
phonon coupling~\cite{sun2} is one of the most interesting properties of II-VI-oxide QWs for optoelectronics. It is necessary from technological viewpoints to correctly describe the well width and composition dependences of the excitonic binding energy.
\begin{figure}[htb]
	\includegraphics[width=.4\textwidth]{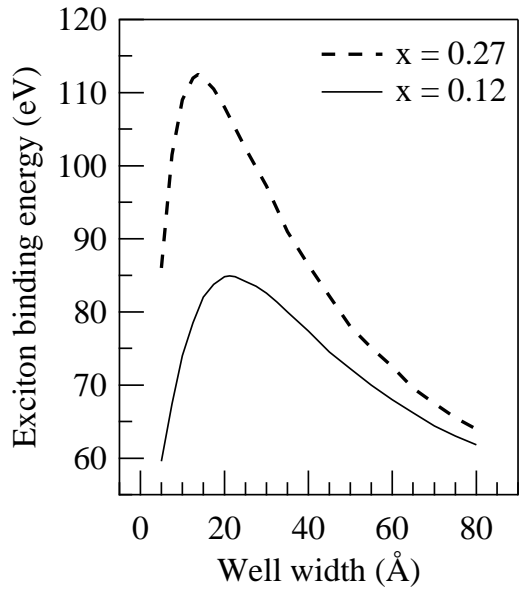}
	\caption{Variational calculation of the exciton binding energy of ZnO/ZnMgO QWs of different composition performed by Coli and Bajaj.}
	\label{calcbindingenr}
\end{figure}

They also calculated the excitonic transition energies in ZnO/MgZnO QW heterostructures, accounting for the effects of the exciton-phonon interaction in the calculation of the exciton binding energies  as formulated by Pollmann and B\"uttner~\cite{pollmann1}. Figure 2 shows their results, describing exciton binding energy as functions of well width and of Mg concentration. Type-I QWs like ZnO/MgZnO exhibit binding energies ranging between the bulk value (about 60~meV) up to $\approx$120~meV or more, depending on the well depth (barrier height) and well width (Fig~\ref{calcbindingenr}). In section \ref{sec:linear} we will compare the calculated excitons binding energies with the available experimental data.

Another important figure of merit of the ZnO-based MQWs is the considerable enhancement of the optical absorption strength. For light propagating perpendicular to the basal plane of the QW, the integrated absorption strength is significantly larger than the
corresponding values of II-VI (ZnSe) or III-V (GaAs) QWs. This finding together with the extremely large exciton binding energy suggests that excitons play a very important role in the linear and nonlinear optical response of ZnO-based QWs. It is therefore very important to have a precise description of the excitonic properties in order to understand some of the relevant optoelectronic phenomena of interest for modern technology, \textit{e.g.}, lasing,
optical modulation and nonlinear switching.

\section{Polarization in wurtzite II-VI-oxides}
\label{sec:QCSE}
A wurtzite structure does not have any symmetry operation that laps the \textit{c}-axis bond parallel to the [0001] directions on the bond of other oblique directions one over the other. Consequently, wurtzites can possess spontaneous polarization without breaking its symmetry. In ZnO like GaN, a double layer coupled in an oblique direction displaces so as to approach its plane directions, which gives rise to the spontaneous polarization along the [000\=1] directions. Furthermore for example, a tensile-stressed situation inside a (0001) plane leads to the generation of piezoelectric polarization field along the [000\=1] directions as a result of piezoelectric effects. Therefore, at a (0001)-heterojunction interface, a sheet charge is present due to these polarizations. Unless formation of defects or reconstruction of the interfaces cancels this sheet charge, these polarization fields give rise to a potential difference at the opposite sides of the well, the magnitude of which is proportional to the well layer thickness ($L_w$).

As is well known, spotaneous polarization and piezoelectric polarization have played a crucial role in the optical properties of nitride-based heterostructures and devices. On the other hand, despite the same crystal structure, it has been believed that these effects are not very serious in the case of ZnO-based QWs, because lattice mismatch between the barrier and well layers is much smaller than that of nitride-based heterostructures. Such an argument is correct as long as contribution from the piezoelectric polarization is discussed.
\begin{figure}[htb]
	\includegraphics[width=.4\textwidth]{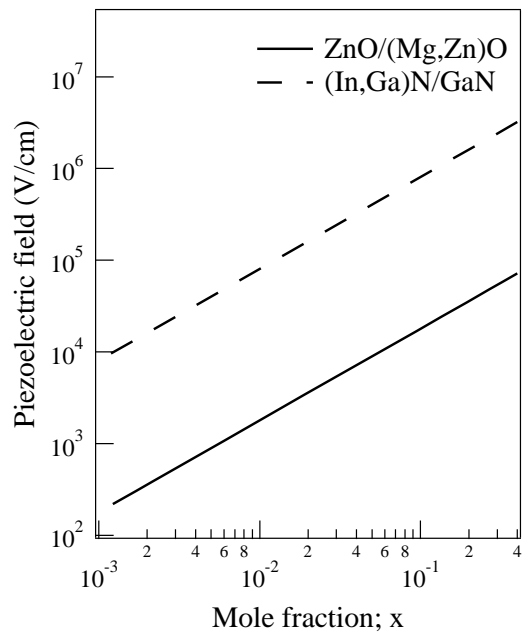}
	\caption{Magnitude of a piezoelectric field by the mismatch-induced strain in QWs as a function of mole fraction, $x$, in In$_x$Ga$_{1-x}$N/GaN (dashed line) and ZnO/Mg$_x$Zn$_{1-x}$O (solid line) QWs. The data for In$_x$Ga$_{1-x}$N/GaN are from Refs.~\protect\cite{schwarz1,goergens1,bykhovski1}.}
	\label{piezo}
\end{figure}

Figure~\ref{piezo} compares the piezoelectric polarization fields as a function of magnesium or indium concentration. The data for In$_x$Ga$_{1-x}$N/GaN are from Refs.~\cite{schwarz1,goergens1,bykhovski1}. On the other hand, we must bear in mind about the possiblity of spontaneous polarization mismatches at the interfaces if the fact that the spontaneous polarization of ZnO is comparable with that of GaN or AlN is considered. Unfortunately, there is no report on the spontaneous polarization coefficient of MgO nor MgZnO alloys. Optical transition energy, oscillator strength, and recombination times crucially depend on the magnitude of polarization fields due to the quantum-confined Stark effects.

\section{Linear optical properties of quasi two-dimensional excitons}\label{sec:linear}
\subsection{Optical absorption}

\begin{figure}[htb]
	\includegraphics[width=.4\textwidth]{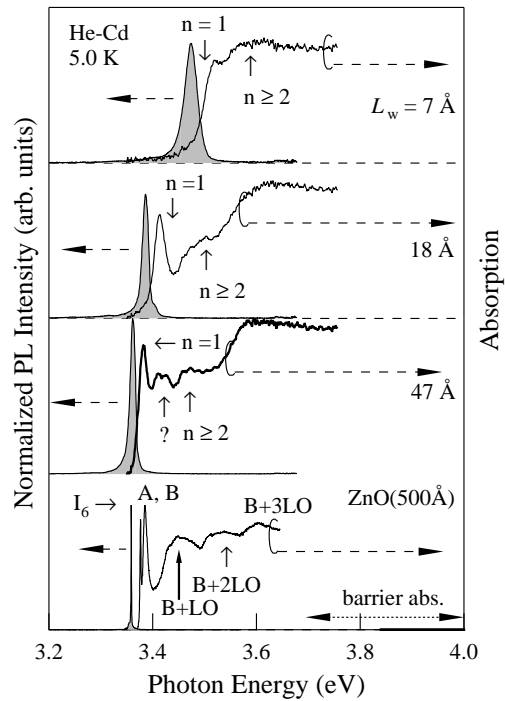}
	\caption{Absorption (continuous lines) and photoluminescence (shadowed spectra) spectra of different ZnO/Mg$_{0.12}$Zn$_{0.88}$O QWs at 5~K. The well widths ($L_{w} $) are: (a) 47~${\rm \AA} $, (b) 17.5~${\rm \AA} $, (c) 6.91~${\rm \AA} $. Absorption energy of barrier layers is shown by a horizontal arrow. Spectra in a 500-\textrm{\AA}-thick ZnO film are also shown. ``A, B'' indicates A- and B-exciton absorption bands, ``I$_{6}$'' shows PL of a bound exciton state, ``B+LO, B+2LO, and B+3LO'' correspond to exciton-phonon complex transitions, ``$n = 1$'' show the lowest excitonic absorption of the well layers, and ``$n \ge 2$'' means the excited states of the exciton or higher interband (sub-band) transitions.}
	\label{MQWT}
\end{figure}

Representative low-temperature absorption spectra of selected ten-period ZnO/Mg$_{0.12}$Zn$_{0.88}$O MQWs are shown in Fig.~\ref{MQWT}. The samples investigated here were grown by laser molecular-beam epitaxy on ScAlMgO$_4$ substrates. Detailed growth procedure has been given
elsewhere~\cite{ohtomo4}. The PL and absorption spectra in a 500-${\rm \AA } $-thick ZnO epilayer on SCAM were included for comparison~\cite{makino7}. The width of these ZnO samples spans the range from bulk-like behavior (47~$\textrm{\AA}$)
to quasi two-dimensional limit (7~$\textrm{\AA}$). The exciton Bohr radius of ZnO is $\approx 18$~\textrm{\AA}~\cite{LBZincoxide}. The combination of these compositional and configurational parameters permits a fine-tuning of the excitonic properties of the ZnO MQWs, which can be investigated systematic optical and structural studies.
Absorption energy of localized excitonic (LO) feature from the barrier layers is shown by a horizontal arrow (3.7--4.0~eV). At lower energies all the samples exhibit distinct excitonic peaks superimposed to the QW continuum. The quantum size effect is clearly demonstrated by the blue shift of the absorption spectrum with decreasing well width. For ZnO-based MQWs, the inhomogeneous broadening becomes comparable to the energies of crystal-field and spin-orbit splittings due to the unavoidable broadening induced when the QWs are constructed. The \textit{A}- and \textit{B}-exciton structures here were not spectrally resolved. The photoluminescence spectra (dashed lines in Fig.~\ref{MQWT}) exhibit clear excitonic bands with typical Stokes shift ranging between 20~meV for wide wells and 40~meV for the narrow wells (cf. Table I of Ref.~\cite{makino11}). These emissions are assigned to excitons localized at the potentials induced by spatial fluctuations on the relevant heterostructure size, because such fluctuation has a more sensitive effect for a very thin well.
\begin{figure}
		\includegraphics[width=.3\textwidth]{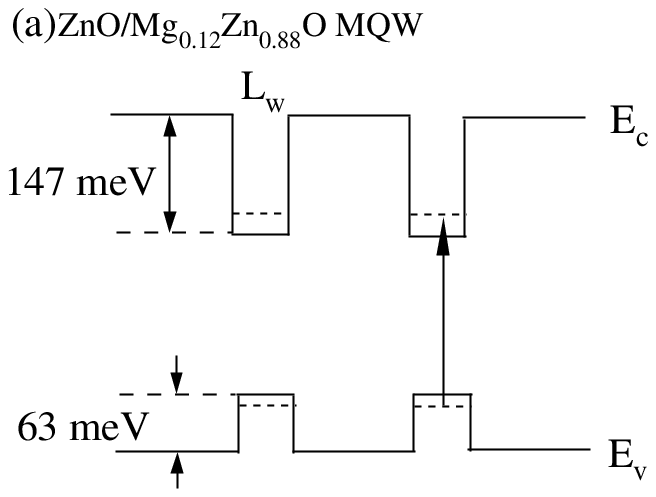}
		\includegraphics[width=.35\textwidth]{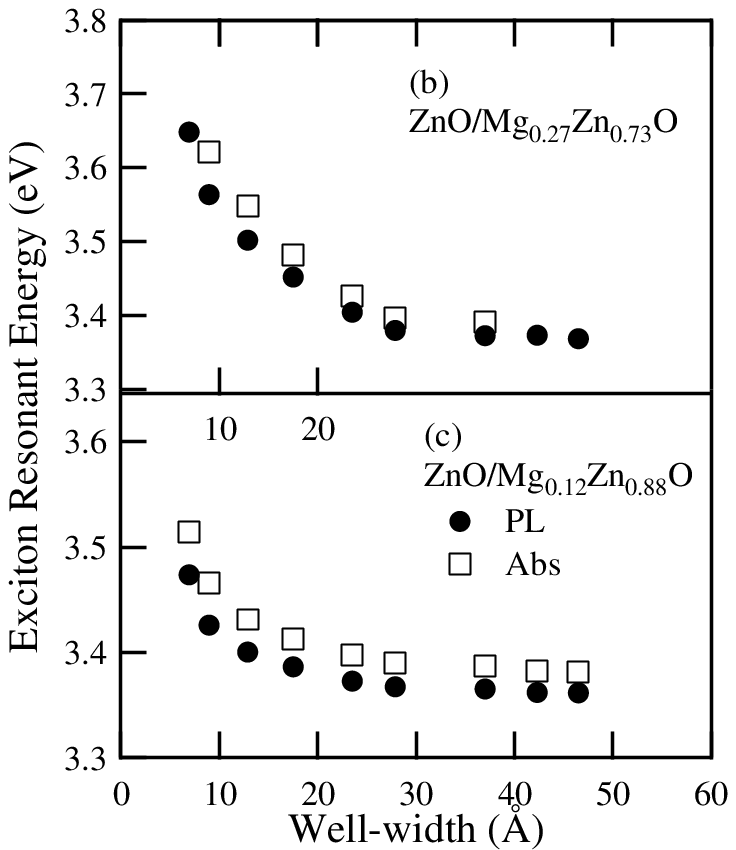}
\caption{(a) Diagram of conduction and valence bands between barrier and well layers in ZnO/Mg$_{0.12}$Zn$_{0.88}$O MQW~\protect\cite{ohtomo4}. The upward arrow shows the lowest interband transition. Peak energies of PL (circles) and absorption (squares) are plotted against $L_{w} $ for ZnO/Mg$_{0.27}$Zn$_{0.73}$O (b) and ZnO/Mg$_{0.12}$Zn$_{0.88}$O (c) QWs. A solid curve is shown as a visual guide.}
\label{MQWPEP}
\end{figure}

Figures~\ref{MQWPEP}(b)--(c) show the well width dependence of the peak energies of PL (closed circles) and absorption (open squares), respectively, in ZnO/Mg$_{0.12}$Zn$_{0.88}$O and ZnO/Mg$_{0.27}$Zn$_{0.73}$O MQWs on SCAM substrates. Energy diagrams of conduction and valence bands in a ZnO/Mg$_{0.12}$Zn$_{0.88}$O MQW are shown in Fig.~\ref{MQWPEP}(a). Changes in composition $x$ cause variation in the depth of the QW. Samples with higher Mg content show absorption spectra extended toward the higher energy side, reflecting the
increased depth of the QW. This result can be explained by considering the fact that confined potential is deeper for a higher barrier. The Stokes shift as well as the width of the PL band in the case of $x = 0.27 $ barrier layers are larger than those of an $x = 0.12 $ barrier with the same $L_{w} $. The depth fluctuation of the potential well is thought to be a reason for this enhancement. Since the $x = 0.27 $ is above the solubility limit, microscopic composition fluctuation is much larger than that in the barrier with $x = 0.12 $. The inhomogeneiety of the band-gaps in the barrier layers induces the depth fluctuation and the enhancement of the exciton localization energy. The unintentional modulation of the QW depth would cause themselves a fluctuation of several meV in the quantization energies of the carriers in the wells.
As described later, sometimes there is an additional PL band below these ``\textit{LE}'' bands.
\begin{figure}[htb]
	\includegraphics[width=.25\textwidth,angle=270]{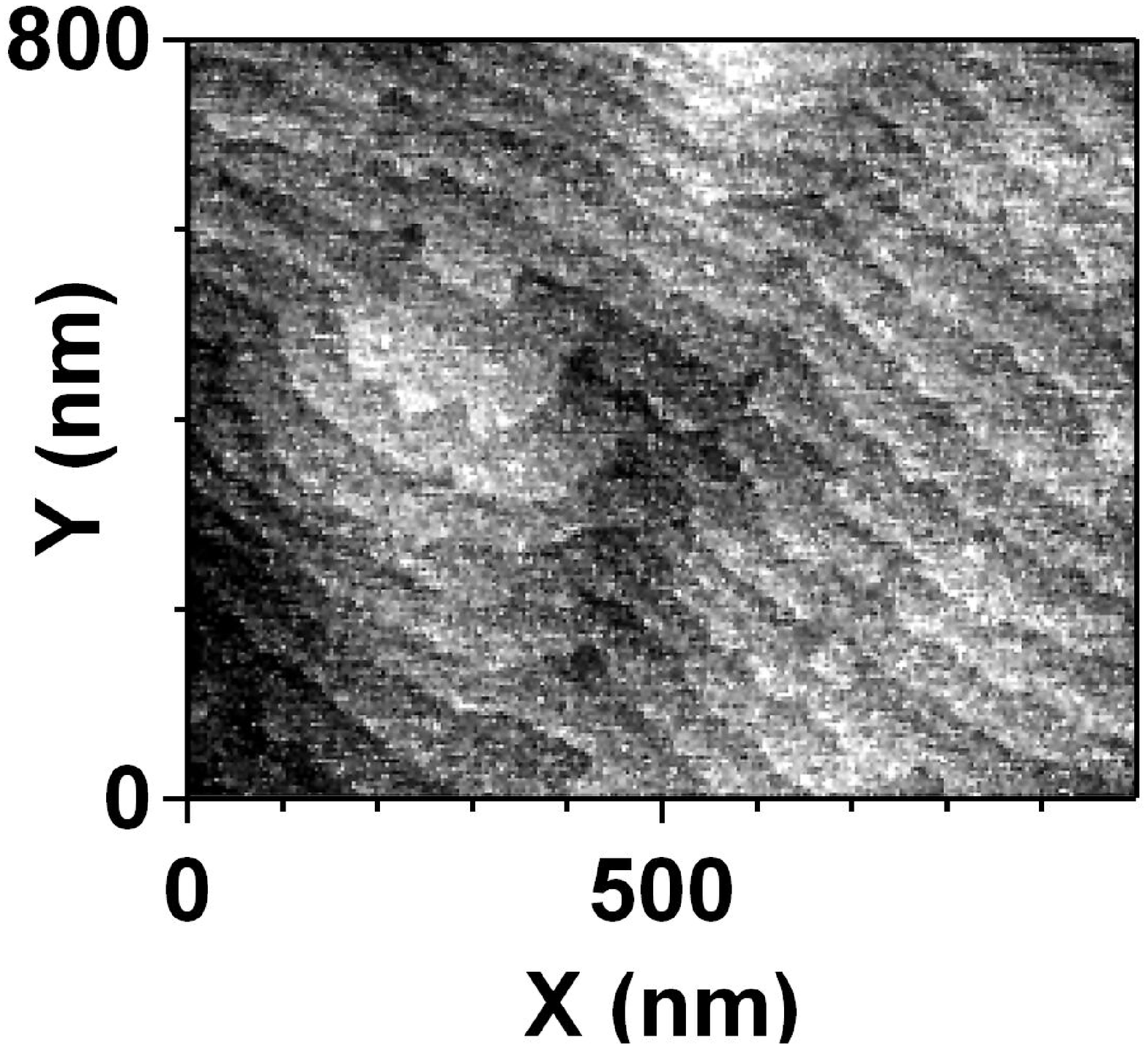}
	\includegraphics[width=.35\textwidth]{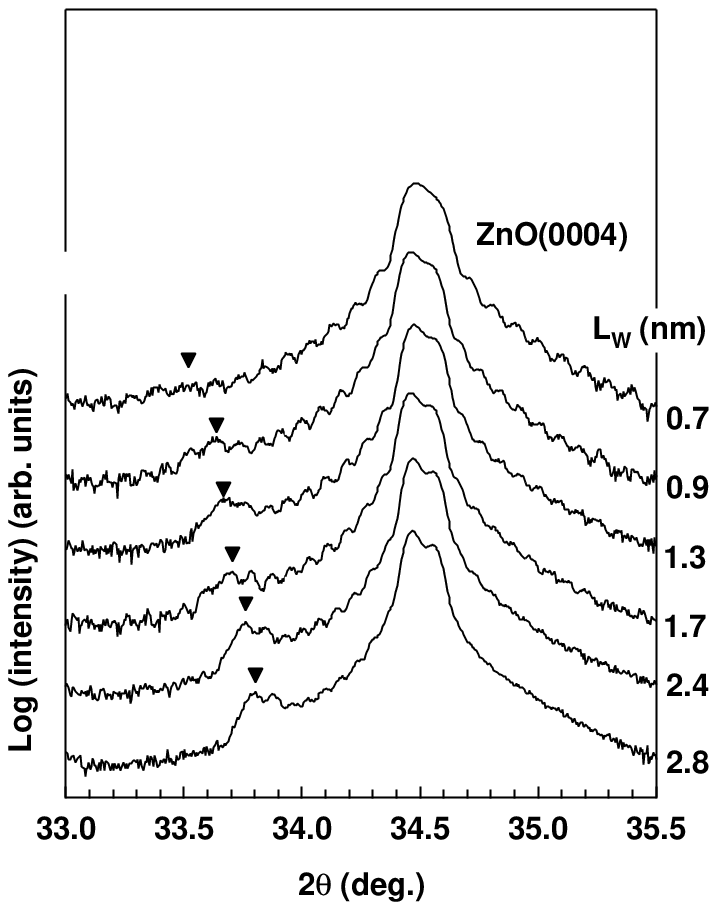}
	\caption{Double-crystal X-ray diffraction pattern of a ZnO/ZnMgO MQW structure having various $L_w$ (28~${\rm \AA} $--6.91~${\rm \AA} $). The barrier width is 50~\AA. The triangles mark the Bragg diffraction peaks corresponding to the superlattice period.}
	\label{XRD}
\end{figure}

X-ray diffraction studies show distinct satellite peaks due to the superlattice periodicity. A set of typical X-ray diffraction patterns for the samples having various thicknesses ranging from 7 to 28~\textrm{\AA} is shown in Fig.~\ref{XRD}.
An analysis of these X-ray patterns provides the information on well width and composition. High crystallinity
and high thickness homogeneity are evidenced by bragg diffraction peaks (closed triangles) and clear intensity oscillations due to the Laue patterns.

A line shape analysis of the satellite peaks reveals some inhomogeneous fluctuation presumably caused by a smooth long-range modulation of the QW thickness. Such an effect produces relevant potential modulation in the QW causing carrier or exciton localization. As stated above, this localization affects dramatically the optical properties of ZnO QWs. For ZnO/ZnMgO QWs, it is unnecessary to
consider the effects~\cite{ohtomo7} of interface diffusion of Mg whereas the formation of ternary alloy at the interface of binary/binary QWs has been experimentally investigated in ZnSe-related QWs~\cite{cingolanirev}.
\begin{figure}[htb]
	\includegraphics[width=.4\textwidth]{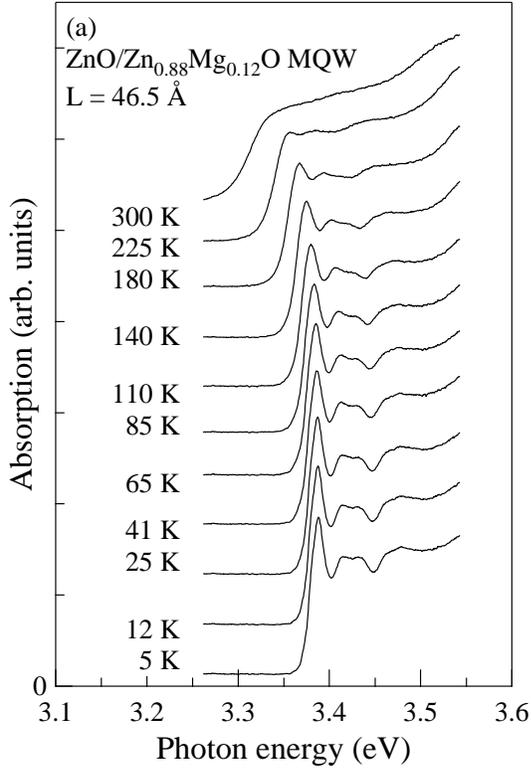}
	\caption{Optical absorption spectra of a ZnO/Mg$_{0.12}$Zn$_{0.88}$O MQW sample with a well width of 47~${\rm \AA} $ at various temperatures, where $E^{Barrier}$ denotes the band-gap energy of the barrier layers. All the spectra have been relatively shifted in the vertical direction for clarity.}
	\label{temp-dep-abs}
\end{figure}

The interactions of electrons with phonons have been demonstrated to greatly affect the optical and electrical properties of semiconductors~\cite{dukeandmahan,rudinandsegall1,pelekanos1}. The temperature dependence of the absorption spectra also provides information on the thermal stability of excitons~\cite{sun2}. In Fig.~\ref{temp-dep-abs}, we show absorption spectra in the 5 to 300-K temperature range for MQWs with $x = 0.12$ and $L_w = 47$~\textrm{\AA}. As can be seen from the figure, even a shallow QW (lower Mg content) exhibit room-temperature excitonic absorption. The thermal stability of the excitons depends on the ratio of the exciton binding energy to the longitudinal optical phonon energy, and on the actual strength of the exciton-phonon coupling~\cite{pelekanos1}. The latter parameter can be approximately estimated from the temperature-dependent absorption linewidth of the ground-level exciton states.
\begin{figure}[htb]
	\includegraphics[width=.33\textwidth]{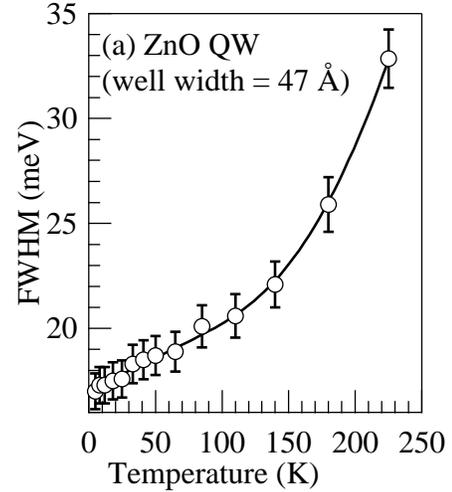}
	\caption{(a) FWHMs of ZnO MQWs with a QW thickness of 47~${\rm \AA} $ as a function of temperature. The solid line represents a result of fit to Eq.~(\protect\ref{e2}). (b) Exciton-phonon coupling constant ($\Gamma_{LO}$) versus the well width $L_w $ at $x = 0.12$ (circles) as obtained from the temperature-dependent exciton width. The solid curve is only a visual guide.}
	\label{fwhm}
\end{figure}

An increase in linewidth of the exciton is observed in the temperature-dependent absorption spectra. Figure~\ref{fwhm}(a) shows the temperature dependence of the FWHM of the excitonic absorption peak for a sample with $L_w = 47$~\textrm{\AA}. Thermal broadening of the excitonic absorption peak is generally interpreted as due to an exciton-phonon interaction. The exciton-phonon interactions ionize the exciton into a free electron and hole in the continuum of states or scatter the exciton into (higher-lying) discrete exciton states~\cite{sun2}. According to Segall's expression~\cite{rudinandsegall1}, the total linewidth of the exciton contains two contributions: inhomogeneous broadening and homogeneous broadening. Within the adiabatic approximation, and then, the temperature dependence of the full width at half maximum (FWHM) can be approximately described by the following equation:

\begin{equation}
    \Gamma(T) = \Gamma_{inh} + \gamma_{ph}T+ \Gamma_{LO}/[\exp(\hbar  \omega _{LO}/k _{B}T) - 1],
   \label{e2}
\end{equation}
where $\Gamma_{inh}   $ is temperature-independent term that denotes the inhomogeneous linewidth due to the exciton-exciton, exciton-carrier interaction~\cite{rudinandsegall1}, and the scattering by defects, impurities and the size fluctuations. The second term $\gamma_{ph} T$ is due to the acoustic phonon scattering. The $\gamma_{ph}$ represents the acoustic phonon coupling strength, mainly caused by the deformation potential mechanism. At low temperatures, since the population of the longitudinal-optical (LO) phonon is vanishingly small, the scattering is mainly ruled by acoustic phonons. The third term is the linewidth due to the LO phonon scattering. $\Gamma_{LO}$ is the exciton-LO phonon coupling strength, and $\hbar \omega_{LO} $ is the LO-phonon energy. At high temperatures, the LO phonon Fr\"ohlich scattering dominates.

The solid line represents the fitted result based on Eq.~(\ref{e2}). The best fit is obtained for the parameter values $\Gamma_{inh}    = 17$~meV, $\gamma_{ph}    = 31$~$\mu $eV/K, and $\Gamma_{LO}   = 341.5$~meV. Here, we take $\hbar \omega_{LO}$ to be 72~meV, equal to that of bulk ZnO. The value of $\hbar \omega_{LO}$ does not exhibit an obvious change of well width for the MQWs as determined by the PL spectra.

We summarize $\Gamma_{LO}  $'s for different QW widths in Fig.~\ref{fwhm}(b). As a comparison, $\Gamma_{LO}$ for a ZnO epitaxial layer was included.
The exciton-phonon coupling in all the QWs ($L_w \le 47$~\textrm{\AA}) assessed here is found to be smaller than in bulk ZnO. In addition, they monotonically decrease with decreasing $L_w$. We assigned this variation to the enhancement of the exciton binding energy. As stated above, $\Gamma_{LO}$ is also dependent on the ratio of the exciton binding energy to the longitudinal optical phonon energy. For bulk ZnO, the exciton binding energy is much smaller than that of LO-phonon. On the other hand, this is no longer the case for ZnO multiple quantum wells (MQWs): the binding energy exceeds the $\hbar \omega_{LO} $. A similar effect was also observed in other QW systems~\cite{pelekanos1}.
The important implication of this result is that the reduced exciton-phonon coupling in QWs favors the exciton stability
leading to a dominant excitonic role in the optical processes of ZnO QWs under strong injection or high temperatures. For comparison, $\Gamma_{LO}$ values for GaAs~\cite{alperovich1}, ZnSe~\cite{fischer2}, CdTe~\cite{LeeJohnsonAustin,rudinandsegall1}, CdS~\cite{rudinandsegall1} and GaN~\cite{xbzhangGaNrev} are 5, 81, 17, 41 and 375--525~meV, respectively. It can be seen that the values of $\Gamma_{LO}$ of ZnO and its QWs are both larger than those in GaAs and even in other II-VI semiconductors. The large value of $\Gamma_{LO}$  suggests that the exciton-LO phonon Fr\"ohlich interaction significantly affects the room temperature and high-temperature performances of ZnO-based devices.

\begin{figure}[htb]
	\includegraphics[width=.4\textwidth]{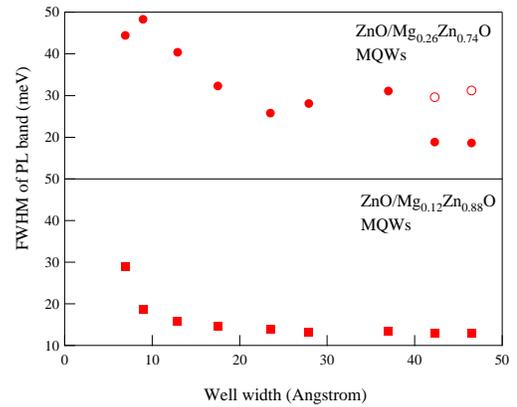}
	\caption{Width of 5-K excitonic PL lines versus $L_w$ in ZnO-based QWs for $x=0.12$ and $x=0.27$. The open circles are estimated from the QCS PL bands.}
	\label{PLwidth}
\end{figure}

Figure~\ref{PLwidth} shows the width of 5-K excitonic PL lines plotted as a function of $L_w$ for comparison.

\begin{figure}[htb]
\includegraphics[width=.33\textwidth]{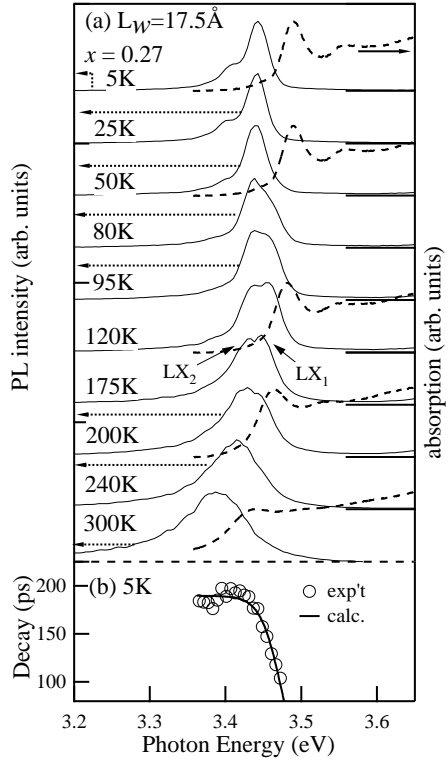}
\caption{(a) PL (solid line) and absorption (broken line) spectra in a ZnO
(17.5~\AA)/Mg$_{0.27}$Zn$_{0.73}$O MQW over the temperature range of 5 to
300~K. All of the spectra have been normalized and shifted in the vertical
direction for clarity. (b) PL decay times as a function of monitored photon
energy at 5~K in the same MQW. The dotted curve is results of the theoretical calculation based on the model of the excitonic localization.}
\label{spectrum-tv}
\end{figure}

It is found that the excitonic luminescence in the ZnO MQWs under investigation
is attributed due to the radiative recombination from the excitons localized
by the fluctuation with spatial well width fluctuation etc.
The evidences of our spectral assignment are; (1) the well width dependence of Stokes shift (energy difference of absorption and luminescence bands), (2) the temperature dependence of PL spectra, and (3) the spectral distribution (luminescence energy dependence) of decay time constants of luminescence.
The typical example of the spectral distribution of the decay time constant is shown
in the lowest curve of the Fig.~\ref{spectrum-tv}.

Here, the temperature dependence of the PL spectrum in a quantum well in case magnesium composition is 0.27 is reported in detail.

\begin{figure}[htb]
\includegraphics[width=.33\textwidth]{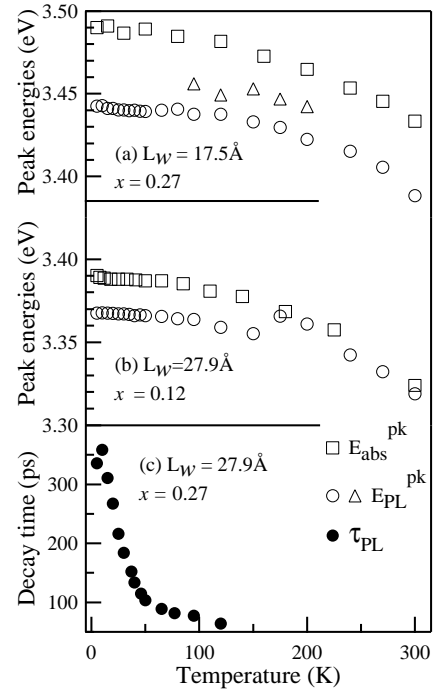}
\caption{PL (solid circles and triangles) and absorption (solid squares)
peak positions as a function of temperature in ZnO(17.5~\AA)/Mg$_{0.27}
$Zn$_{0.73}$O (a) and ZnO(27.9~\AA)/Mg$_{0.12}$Zn$_{0.88}$O (b) MQWs.
(c) Temperature dependence of PL decay times, $\tau_{PL}$ in
ZnO(27.9~\AA)/Mg$_{0.27} $Zn$_{0.73}$O MQWs at 5--120~K~\protect\cite{makino11}.}
\label{fig:pkplot}
\end{figure}

Figure~\ref{spectrum-tv}(a) shows temperature dependence of PL (solid line) and
absorption (broken line) spectra in ZnO(17.5~\AA)/Mg$_{0.27}$Zn$_{0.73}$O
MQWs over a temperature ($T$) range from 5--300~K. One should bear in mind that
spectra taken at
95 to 200~K encompassed two peaks, both of which originated from
recombination of localized excitons. The separation of these peaks was
12 to 20~meV. Figure~\ref{spectrum-tv}(b) shows PL decay times as a function of monitored photon
energy at 5~K in the same MQW. The dotted curve is results of the theoretical calculation based on the model of the excitonic localization. Figure~\ref{fig:pkplot}(a) summarizes peak energies of the PL spectra ($E_{PL}^{\text{pk}}$) (solid circles and triangles) and the
excitonic absorption energy (solid squares) as a functions of temperature. It should be noted that the higher PL peak position does
not coincide with that of absorption spectra even at the temperatures approximately equal to room temperature. 

We also examined, for comparison, the temperature dependence of PL peak
energy in an MQW having a lower barrier height: a ZnO/Mg$_{0.12}$Zn$_{0.88}$O
MQW with a well width of 27.9~\AA. Figure~\ref{fig:pkplot}(b) shows the peak
energies of PL (circles) and absorption (squares) spectra in this sample.
In this case, contrastingly, the energies of luminescence and absorption coincide each other at the temperatures near room temperature.
Two kinds of MQWs having different barrier
heights showed significantly
different temperature dependences of PL spectra.

Followed on a temperature rise, the PL energy of ZnO(17.5~\AA)/Mg$_{0.27}$Zn$_{0.73}$O
MQWs exhibited low energy shift between 5K and 50K, exhibited the higher energy shift between 50 and 200~K, and again shifts to a low energy side at the temperatures more than 200~K.

Furthermore, between 95 and 200~K, the spectra encompassed two peaks, both of which originated from
recombination of localized excitons.
The absorption peak energies both in ZnO epilayers and in MQWs are
monotonically decreasing functions of temperature, as revealed in
previous studies. This is attributed to the temperature-induced shrinkage of fundamental energy gap.

In general, when a dominant PL
peak is assigned to radiative recombination of localized excitons, its peak
energy blueshifts with increasing temperature at low temperatures and
redshifts at higher temperatures. The $E_{PL}^{\text{pk}}$ blueshifts and
continuously connects to that of free excitons due to thermal activation of
localized excitons. The $E_{PL}^{\text{pk}}$ of the free-excitonic emission
is a monotonically decreasing function of temperature due to the bandgap
shrinkage. The temperature dependence shown in Fig.~\ref{spectrum-tv}(a) is,
however, different from the abovementioned typical behavior.

The temperature dependence of the recombination mechanism for the localized
excitons is considered to closely relate to the temperature variation
of the decay time constant of their PL. Thus, the temperature dependence of PL decay times
($\tau_{PL}$) in the MQW with well width of
27.9~\textrm{\AA} is shown in Fig.~\ref{fig:pkplot}(c). The $\tau_{PL}$ values
exhibites a nonmonotonical behavior with respect to the temperature; it increased in the low tempearature range, while it decreased above a certain critical temperature.

The temperature dependence of the recombination mechanism for the localized
excitons can be explained as follows: (i) For 5~K
\textless $T $\textless 50~K, the relatively long relaxation time of
excitons gives the excitons more opportunity to relax down into lower
energy tail states caused by the inhomogeneous potential fluctuations
before recombining. This is because the radiative recombination
processes are dominant compared with nonradiative processes in this
temperature range. This behavior produces a redshift in the peak energy
position with increasing temperature. (ii) For 50~K \textless $T
$\textless 95~K, the exciton lifetimes decrease with increasing
temperature. Thus, these excitons recombine before reaching the lower
energy tail states. This behavior enhances a broadening of the higher-energy
side emission and leads to a blueshift in the peak energy. (iii) For 95~K
\textless $T $\textless 200~K, further enhancement of high-energy emission
components produces a new peak, as seen in Fig.~\ref{fig:pkplot}(a)
(triangles). (iv) Above 200~K, since the
excitons are less affected by the temperature-induced rapid change in their
lifetime and relaxation rate are increased due to the increased phonon
population, blueshift behavior therefore becomes less pronounced. 

Since
the energy of blueshift is smaller than the temperature-induced bandgap
shrinkage, the peak position again exhibits a redshift behavior. As
mentioned above, the features for excitonic spontaneous emission in the
well layers are sensitively affected by the dynamics of the
recombination of the localized exciton states, which significantly vary with temperature.

\begin{figure}[htb]
	\includegraphics[width=.4\textwidth]{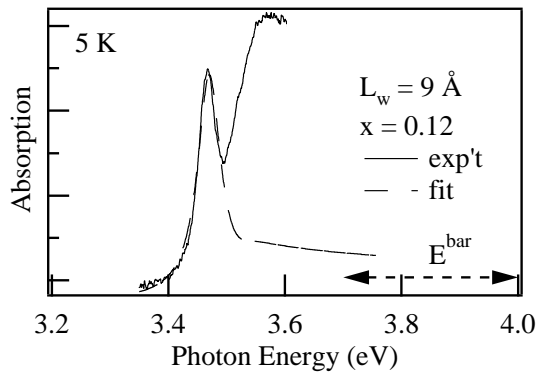}
	\caption{Line-shape fitting of the absorption spectrum at 5~K of a ZnO/ZnMgO MQW with $L_w = 9$~\AA. The simulation result (a dashed line) and the experimental data (a continuous line) are shown. Note that the A- and B-exciton structures here were not spectrally resolved.}
	\label{lineshape-fit}
\end{figure}

In order to get information on this important parameter and more in general on the excitonic eigenstates and binding energy one has to model the absorption line-shape with the modified Elliot two-dimensional (2D) exciton model~\cite{goni1,gurioli1}. It is well known that the reduction of the dimension
modifies the excitonic line shape which are characterized by the enhancement of excitonic
binding energies and by the concentration of the density-of states
into the discrete excitonic peak. For purely two-dimensional excitons this
has been first calculated by Shinada and Sugano~\cite{shinada1}, Such a purely 2D formulation has been used in this work for
simplicity. 

Analytical expression including the broadening effect has been deduced in Refs.~\onlinecite{goni1,gurioli1} and is used here for the simulation. For example, the step like continuum of the QW density of states is simulated by a step function, convoluted with a lorentzian broadening used to reproduce the exciton factor at the band edge.
An example of the result of this line-shape simulation is shown in Fig.~\ref{lineshape-fit} for a ZnO/ZnO/Mg$_{0.27}$Zn$_{0.73}$O MQW with $L_w=9$~\textrm{\AA}. The parameters used in the simulation are the band gap energy of $E_g$, the excitonic binding energy of $E^b_{Ex}$, and the damping parameter of of $\Gamma$. The value of $E_g$ was 3.54~eV deduced from the lowest excitonic line, whereas the $E^b_{Ex}$ of 72~meV was cited from the results of theoretical work~\cite{coli1}. The damping parameter
of 24--25~meV was deduced from the width of excitonic absorption peak. The calculated absorption spectrum (a dashed line) reproduces well the experimental spectrum (a solid line) for the lineshape of the lowest excitonic absorption, whereas above the excitonic line, the calculation is not in good agreement with the experimental data. There are two plausible reasons for the poor agreement. Firstly, the tailing contribution of the localized excitonic states of the barrier (MgZnO layers) should be pointed out.

It is, however, difficult to quantatively estimate the contribution of the barrier absorption. Secondly, the approximation of pure two-dimensional excitons is also responsible for that: one has to
know quantatively the realistic degree of anisotropy of our QW (i.e., the sommerfeld factor). However,
unfortunately in ZnO QWs these parameters has not been yet known.

\begin{figure}[htb]
	\includegraphics[width=.33\textwidth]{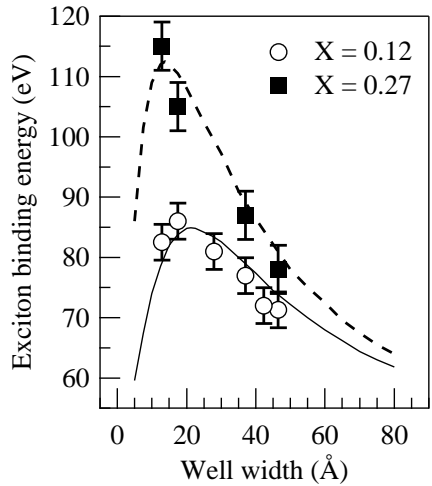}
		\caption{Experimental exciton binding energy of ZnO/ZnMgO QWs with $x=0.12$ (circles) and $x=0.27$ (triangles) as a function of the well width. The curves (solid and dashed lines) are results of variation calculation~\protect\cite{coli1}.}
	\label{bindingenergy}
\end{figure}

Because of the abovementioned disagreement, the experimental exciton
binding energy is extracted from the energy positions of the
stimulated emission as a result of exciton-exciton scattering and
are summarized in Fig.~\ref{bindingenergy} for ten different MQWs
(open circles and closed squares for $x=0.12$ and $x=0.27$ QWs,
respectively), together with the results of the variational
calculations (solid and
dashed lines) described in section~\ref{sec:excitontheory}. The method of evaluation will be described later in
section~\ref{sec:nonlinear}. For shallower MQWs, the calculated
exciton binding energy varies approximately from the bulk value
(about 70~meV for $L_w = 47$~\textrm{\AA}) up to 86~meV for $L_w= 18$~\textrm{\AA}. 

For even lower well width, the calculated
exciton binding energies decreases due to the increased penetration
of the exciton wavefunction in the barriers. Deeper QWs
($x=0.27$) exhibit a similar behavior, but the maximum exciton
binding energy reported is 115~meV. The experimental results are in
good agreement with the theoretical calculation. The calculated exciton binding energy match within 5~meV the experimental values in the case of the
$x=0.12$ MQW (open circles and continuous line). The reasonable overall agreement between calculated and measured binding energies supports the choice of polaron masses and dielectric constants used here.

These results clearly show that exciton confinement in ZnO/ZnMgO MQWs can be tuned independently by varying the well composition ($x$) or thickness ($L_w$). Based on the experimental data, for a given well width of 37~\textrm{\AA}, the increase of the well width (the increase of the Mg content from typically $x=0.12$ to $x=0.27$) results in a 14\% enhancement of the exciton binding energy, reflecting the enhanced localization of the carrier wavefunctions in the well.

\begin{figure}[htb]
	\includegraphics[width=.4\textwidth]{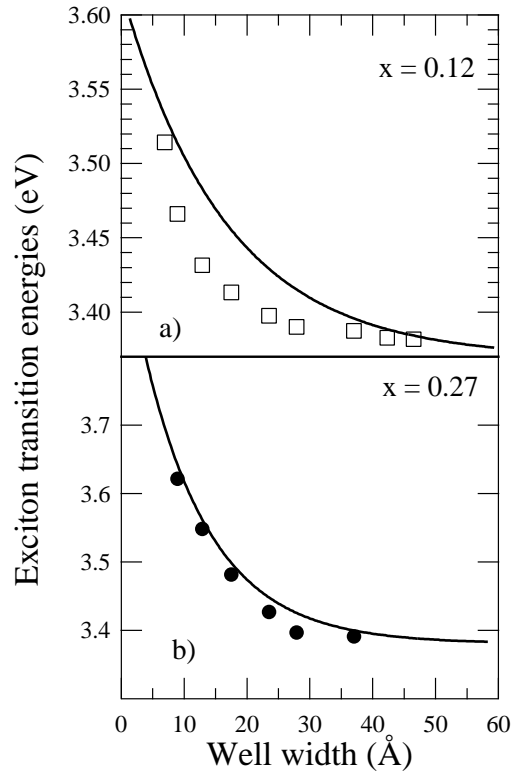}
\caption{Variations of the exciton transition and binding
energies (inset) in ZnO/Mg$_{0.12}$Zn$_{0.88}$O and ZnO/Mg$_{0.27}$Zn$_{0.73}$O MQWs as functions of well width. Symbols refer to
excitonic transition energies measured (squares and circles) by Makino \textit{et~al.} (Ref.~\protect\onlinecite{makino11}). Lines refer to the
results of the calculations in which Coli and Bajaj described the
electron-hole interaction accounting for the exciton-phonon
interaction through Pollmmann-B\"uttner potential (solid and
dashed lines).}
	\label{calcresonen}
\end{figure}

In Fig.~\ref{calcresonen}, we compare calculated exciton resonance
energies with the experimental values (closed circles and open squares). We should emphasize again the importance of
exciton-phonon interaction in these polar semiconductor QWs, because
the better agreement between the results of their calculations (lines) and experimental data if the calculation takes that effect into account.

Before concluding this sub-section we discuss the lineshape of photoluminescence excitation (PLE) spectra taken for an MQW grown on a sapphire substrate. There has been few experimental report of the PLE spectroscopy because the lattice-matched ScAlMgO$_4$ substrate is transparent for the wavelength region of interest. In a QW of the weak confinement regime such as the ZnSe/ZnS QW, strong hot
exciton features are observed in the PLE spectra. Such an oscillatory behavior
with a period equal to the LO phonon energy has been often observed in bulk semiconductors. The oscillatory structure was not observed in the PLE spectrum taken for a ZnO MQW~\cite{ohtomo4}, indicating a significant reduction of the exciton-phonon coupling (the Huang-Rhys factor) with respect to bulk ZnO.

\subsection{Temporal evolution of the excitonic transitions}

In this subsection we discuss the transient properties of excitons confined in ZnO-based
QWs. Only time-resolved PL experiments have been performed for ZnO/Zn$_{0.88}$Mg$_{0.12}$O MQWs in which it is unnecessary to take the
internal electric field effects into account. In optical experiments, excitons are formed with some excess energy by the off-resonant pumping. The excess energy is relaxed in a few hundreds fs by the exciton LO phonon interaction, resulting in a quasi-equilibrium distribution of thermalized excitons. After this short transient  excitons undergo a number of different interactions, namely localization at potential fluctuations, scattering with electrons or other excitons and eventually recombine radiatively on a time scale of the order of few hundred ps.

\begin{figure}[htb]
	\includegraphics[width=.4\textwidth]{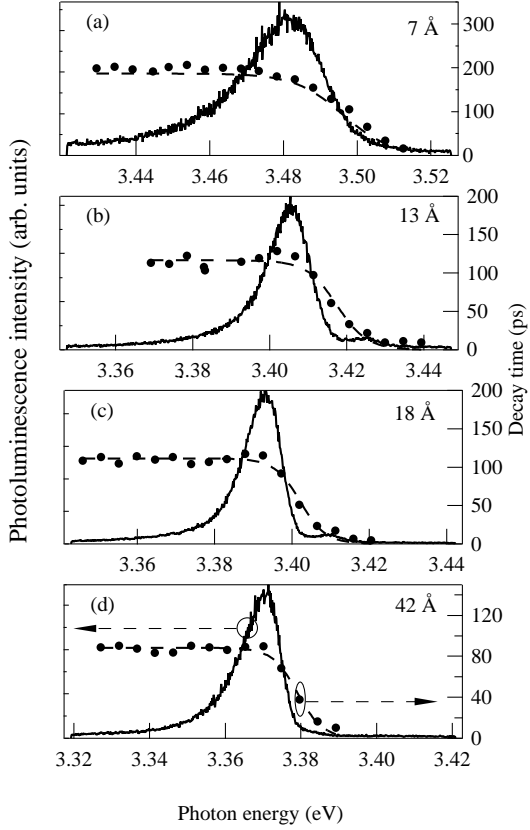}
	\caption{Time-integrated PL spectra (solid traces) and spectrally-resolved PL decay times (closed circles) taken at 5~K for four MQWs with $L_w$ of 7 (a), 13 (b), 18 (c), and 42~\textrm{\AA} (d) are shown. The dashed curves are the theoretical ones, indicating that the increase of the decay time occurring in the low energy tail of the exciton band due to trapping and localization.}
	\label{TRPL}
\end{figure}

On the time scale of the exciton lifetime the behavior of the ZnO QWs does not differ appreciably from that of other II-VI or III-V structures. In Fig.~\ref{TRPL}, we display the spectral distribution of decay time in ZnO/ZnMgO MQW structures of well width 7, 13, 18, 42~\textrm{\AA} and Mg content $x=0.12$.

The time-resolved PL experiments provide important information on the localization of excitons within the inhomogeneously broadened density of states caused by monolayer fluctuation and disorder in the ternary alloy barriers. As a general trend the temporal evolution of the luminescence from localized excitons exhibits decay time longer than the free excitons. This is clearly seen in the low energy tail of the exciton resonance of ZnO/ZnMgO MQWs. Estimated PL decay time is a monotonically decreasing function of the emission energy. This is because the decay of the localized excitons is not only due to radiative recombination but also due to the transfer process to the tail state. An analysis of the time-resolved luminescence traces provides the density of localization centers existing in the QW. If the density of the tail state is approximated as $\exp (E/E_0)$, and if the lifetime of localized excitons ($\tau_{PL}$) does not change with their emission energy, the observed decay times $\tau (E)$ can be expressed by the following equation~\cite{gourdon1}:
\begin{eqnarray}
\tau(E) = \frac{\tau_{PL}}{\exp ((E - E_{me})/ E_0)},
\label{gourdonformula}
\end{eqnarray}
where $E_0$ shows the degree of the depth in the tail state and $E_{me}$ is the characteristic energy representing the absorption edge. Results of best fits have been given in Ref.~\onlinecite{chia2}. It is found that the evaluated $\tau_{PL}$ is a monotonically decreasing function of $L_w$. Moreover, the $L_w$ dependences of the localization depth [$E_0$ in Eq.~(\ref{gourdonformula})] of excitons and the $\tau_{PL}$ were similar with respect to each other (cf. Fig~3 of Ref.~\onlinecite{chia2}). Therefore, the $L_w$ dependence of $\tau_{PL}$ can be explained as being due to the thermal release effect from localized to delocalized states of excitons.

\begin{figure}[htbp]
	\includegraphics[width=.4\textwidth]{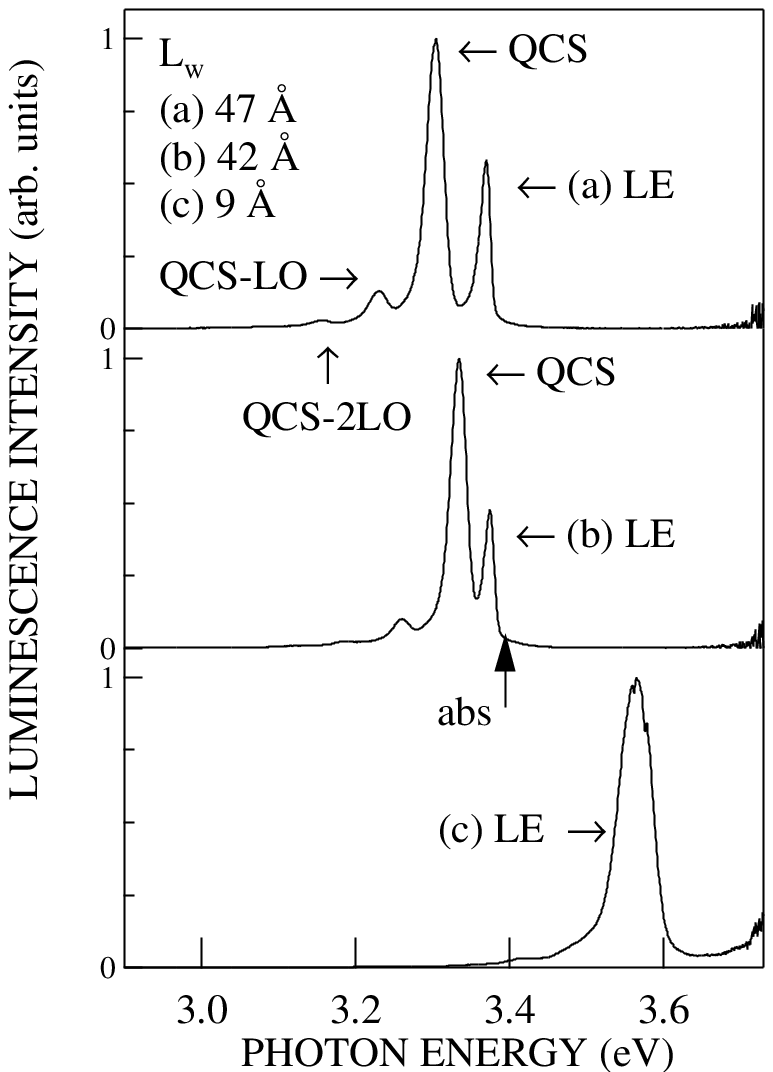}
	\includegraphics[width=.275\textwidth]{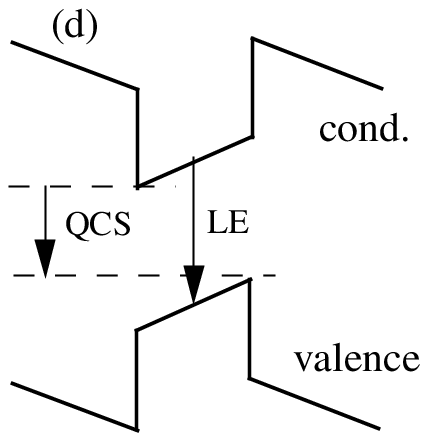}
\caption{5-K PL spectra of ZnO/Mg$_{0.27}$Zn$_{0.73}$O QWs. Spectra (a), (b), and (c) were measured from three samples with different well-widths of the QW samples. Nomenclatures of ``\textit{LE}'' and ``\textit{QCS}'', respectively, mean the localized exciton and the quantum-confined Stark (QCS) effect. The LO-phonon replicas are indicated as arrows in the figure. The energy position of the excitonic absorption theoretically calculated by Coli and Bajaj~\cite{coli1} is shown by the arrow with ``Abs''. (d) Schematic diagram of the conduction and the valence bands of wurtzitic QWs under the spontaneous and piezoelectric polarization field. Electrons and holes distributed in the well regions are spatially separated due to the quantum confined Stark effects. The downward arrow with ``\textit{QCS}'' corresponds to the radiative recombination process influenced by these effects, while the arrow with ``\textit{LE}'' corresponds to that process without their influence. \label{fes}}
\end{figure}

The quantum efficiency suppression could be avoided even for the small $L_w$s below 10~\textrm{\AA}, the reason of which is not clear now nevertheless is highly desirable for UV light-emitter device applications.

\subsection{Quantum-confined Stark effects observed
in ZnO quantum wells}

It is not necessary to take the quantum-confinement Stark effect into account in the case of the ZnO/Mg$_{0.12}$Zn$_{0.88}$O MQWs. In the case of $x=0.27$ in which lattice-mismatch between the well and the Mg$_{x}$Zn$_{1-x}$O barrier layers is relatively large, since the directions of spontaneous and piezoelectric polarizations along the ZnO wells are coincident with respect to each other, the electric-field induced inclination of the band profiles is considered to become significant.

Figure~\ref{fes} shows 5-K PL spectra corresponding to ZnO/Mg$_{0.27}$Zn$_{0.73}$O QW samples with three different $L_w$s. Only the localized exciton (``\textit{LE}'') band was observed in ZnO/Mg$_{0.27}$Zn$_{0.73}$O QWs with $L_w$ smaller than 38~\textrm{\AA }. Increasing a $L_w$ yielded a different observation: Other than zero-phonon peaks of the ``\textit{LE}'' bands, now there is another prominent PL peak (denoted by ``QCS'') in the 47- and 42-\textrm{\AA }-thick QWs, which seem to have difference origins with the ``LE'' band. These ``\textit{QCS}'' bands have larger Stokes shift than that of ``\textit{LE}'' bands. These PL bands are located $\approx 40 $~meV in energy below the emission band of the localized excitons and $\approx 60$~meV below the absorption energy of the free exciton transition. Calculation absorption energies~\cite{coli1} are shown by arrows in Fig.~\ref{fes}.

We think that the magnitude of the electric field in the case of $x = 0.27$ is larger than that of $x = 0.12$, because of the larger lattice-mismatch between ZnO and Mg$_{0.27}$Zn$_{0.73}$O (cf. Fig.~\ref{piezo} in section~\ref{sec:QCSE}). This internal field, present along the growth axis of the system, is caused by piezoelectric and spontaneous polarizations. It is considered to be easier to observe the ``\textit{QCS}'' bands in the sample with higher Mg concentration. Therefore, this band is attributed to be due to the radiative recombination from the excitons influenced by the internal electric field. The Stokes-like shift of PL is due to the quantum-confined Stark effect induced by the internal electric field. If there is a sizable polarization field inside a QW with small lattice mismatch between barrier and well layers, the spontaneous polazation mismatch at interfaces is more likely to be responsible. We cannot however discuss this contribution because the spontaneous polarization coefficient of MgO has not been reported so far.

Why did the QWs with $L_w$ of 7--38~\textrm{\AA } not show the ``\textit{QCS}'' emission? A band diagram of QWs under both the piezoelectric and spontaneous polarization fields is schematically shown in Fig.~\ref{fes}(d). The PL energy of the ``\textit{QCS}'' bands is lower than that of the ``\textit{LE}''. Nevertheless, in the case of small $L_w$, the energy difference can be neglected because the depth of triangle-shaped potential well is smaller than the case of larger $L_w$. Both the electron and hole wavefunctions are confined in the wells even when the electric field is present. Thus, the overall Stokes-like shift of the PL is thus determined only by the in-plane (lateral) band gap inhomogeneity. We observe the single PL peak (``\textit{LE}'' band) in this case. On the other hand, in the opposite case (larger $L_w$s), the depth of triangle-shaped potential well become larger. Carrier wavefunctions drops into these triangle-potentials at one side of the well layer. The electric field pushes the electron and the hole towards opposite sides of the well. Thus, the energy difference between ``\textit{QCS}'' and ``\textit{LE}'' becomes sizable ($\approx 40$~meV).

It is now known that the quantum confined Stark effects somewhat reduce the excitonic oscillator strength. The distinct excitonic peak could not be observed in a 5-K absorption spectrum taken for a 42-\textrm{\AA }-thick QW. However, this is possible for an 18-\textrm{\AA }-thick QW.
Such disappearance may be explained by the oscillator strength quenching. It has been demonstrated similarly to the GaN that the magnitude of piezoelectricity is larger than zincblende semiconductors and that there are effects of spontaneous polarization that are absent in zincblende semiconductors. More detailed theoretical work is desired to determine the magnitude of electric field.

\subsection{LO phonon replica in ZnO/Mg$_{0.27}$Zn$_{0.73}$O QWs}

An important result of the exciton-phonon interactions in semiconductors is the appearance of phonon-assisted emissions of excitons in the luminescence spectra. In the past decades, phonon replicas have been observed in the PL spectra in most II-VI semiconductors~\cite{segallandmahan,gross2,weiher1} and in some ionic crystals~\cite{rudinandsegall1}. The intensities of these replicas relative to the zero-phonon peak depend strongly not only on the exciton-phonon coupling strength but also on the internal electric field in the QW layers. We have already discussed the latter effect in Section~VC. The distribution of emission intensities between phonon replicas and the main emission peak reflects the coupling strength with the
LO phonons and is described in terms of the Huang-Rhys factor.

We give another supporting evidence of our spectral assignment for ``QCS'' bands
by paying attention to the intensity distribution
of LO phonon replicas. The 1LO and 2LO phonon replicas of the ``\textit{QCS}'' bands (e.g., ``\textit{QCS-LO}'')
are clearly seen in Figs.~\ref{fes}(a) and (b). This is not the case for the
localized exciton emission as shown in Fig.~\ref{fes}(c). The $L_w$ dependence of the PL intensity ratio has been given elsewhere~\cite{makino20,makino24}.
The coupling strength of the electron-hole pairs separated due to
quantum-confined Stark effect (``\textit{QCS}'') with the LO phonons is significantly larger
than that of the localized excitons (``\textit{LE}'').

In general, the phonon coupling strength generally depends strongly on the spatial distributions of electron and hole charge densities and sometimes deviates from the bulk value~\cite{kalliakos1}. The deviation is more outstanding in the cases that the wurtzitic heterostructures are influenced by the strong internal electric field. As shown in Fig.~\ref{fes}(d), the electrons and holes are distributed on opposite sides along $c$-axis direction. The distance between them are not only determined by the Coulomb force but also by the electric field. The reduced overlap of these electron and hole charge densities must be responsible for the observed increase of the coupling strength. It is known that, according to Hopfield~\cite{hopfield1}, the coupling strength is a growing function of the distance between electrons and holes. Conversely, this enhancement supports our spectral assignment concerning the ``\textit{QCS}'' band.

\section{Nonlinear excitonic properties}
\label{sec:nonlinear}
In this section we discuss nonlinear optical properties of excitons in ZnO-based QWs. Due to the enhanced excitonic stability, nonlinear optical processes involving excitons are very important in those
QWs and must be considered relevant even for the design of optoelectronic devices operating even with high carrier density and at room temperature. We deal with high-density exciton effects, in which bound electron-hole pair states participate, i.e., the regime of moderately high-excitation intensity is described.

\subsection{Basic theoretical concepts}
\label{basictheoretical}
In this subsection, we give a treatment of the two representative recombination processes in dense exciton systems, which play a crucial role for the excitonic nonlinearlities of the ZnO epitaxial layers and ZnO-based QWs. We discuss uniquely the following two optical processes caused by the interaction between excitons: (1) exciton-exciton (x-x) recombination due to their inelastice scattering and (2) biexciton recombination. In these processes only parts of the energy of the recombining exciton leave the QW as a photon, while the remaining energy is transfered to other excitations, such as free carriers or excitons. These processes have first been introduced in 1968 for a CdS crystal~\cite{guillaume1}.
\subsubsection{Inelastic scattering processes of excitons}
In the x-x recombination, an exciton is scattered converting into a photon, whereas another exciton is scattered into a state with higher energy (\textit{e.g.}, excited-states excitons or continuum states). Therefore, both energy and momentum are conserved in the total collisional process. The x-x processes is a dominant mechanism leading to stimulated emission in ZnO QWs and has been observed at temperatures from 5~K up to well above room temperature. The emission line
of this assignment has been called \textit{P}$_{n}$ ($n$ denoting the quantum
number of exciton).

\subsubsection{Biexcitons}
Since excitons in semiconductors can be understood in analogy to the hydrogen atom, one can expect that excitonic molecules or biexcitons may exist in the same way as H$_2$ molecules. Variational calculations for the biexciton binding energy have been performed as a function of the mass ratio. They give a lower limit of ratio in exciton and biexciton binding energies and indicate that the biexciton is bound in all semiconductors.
In the biexciton recombination processes, a biexcitonic molecule decays into a photon and into a transverse exciton or exciton-like polariton on the lower branch, respectively. The well-width dependence of biexciton binding energy has been estimated experimentally by the low-temperature pump-and-probe spectroscopy. At elevated levels of excitation, stimulated emission related to the presence of a dense neutral electron-hole plasma has been observed. Such very high-excitation regime will be discussed later.

\subsection{Results for the emission due to the inelastic scattering processes: role of excitons in the stimulated emission}
\label{sec:biexciton}

\begin{figure}[htb]
	\includegraphics[width=.4\textwidth]{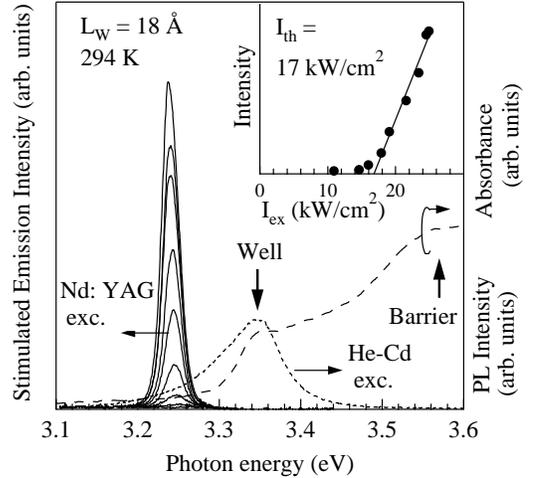}
\caption{Excitation intensity ($I_{ex}$) dependence of the stimulated
emission spectra in a ZnO/Mg$_{0.12}$Zn$_{0.88}$O SL ($L_{w}$=18~\textrm{\AA}) under
pulsed excitation measured at room temperature. Spontaneous PL (a dotted line) under continuous-wave
excitation and absorption (a broken line) spectra are also shown. Inset
depicts the integrated intensity of the stimulated emission peak as a
function of $I_{ex}$. Threshold intensity ($I_{th}$) is 17~kW/cm$^{2}$.}
\label{rawdata}
\end{figure}

In this section we start from the experimental results that are related to the x-x recombination process in order to discuss the interplay between excitons recombination in the stimulated emission mechanism of ZnO MQWs. A method to determine the exciton binding energy from this x-x emission will be explained. Because the excitons of ZnO are stable even at room temperature, the x-x inelastic scattering is expected to be seen in the wide temperature range in their MQWs. This process has been observed at intermediate excitation levels in all II-VI compounds investigated so far. The exciton-exciton process can be distinguished spectroscopically by the characteristic energy of their luminescence or stimulated emission. In general, energy and momentum conservation and Boltzman distribution for the excitons are assumed in the scattering process. For example, typical emission band is displaced from the position of the free exciton to lower energies by approximately one exciton binding energy ($E_{x}^{b}$). Since we would like to discuss its temperature dependence, the resulting emission energy for the exciton-exciton scattering should be written as a function of temperature for the 2D case:
\begin{equation}
\hbar \omega_{\rm x-x} \simeq E_{HH} - E_b - 2 \delta k_B T,
\label{kinetic}
\end{equation}
with $0 \ll \delta \le 1$ where $E_{HH}$ is the free exciton (heavy hole) energy, and $\mu $ is the reduced exciton mass. Typical room-temperature spectra in MQWs with $x=0.12$
and $L_{w}=18$~\textrm{$\AA$} are shown in Fig.~\ref{rawdata}. This observation corresponds to the demonstration of room-temperature excitonic gain. The \textit{P}$_n$ lines are observed for the MQWs. Strong and sharp emission peaks were observed at 3.24~eV above
a very low threshold ($I_{th}$=17~kW/cm$^{2}$), and their integrated
intensities rapidly increased as the excitation intensity ($I_{ex}$) increased, as can be seen in the inset. Unlike in the case of epitaxial ZnO thin films, a fine structure associated with the $P$-bands (\textit{P}$_2$, \textit{P}$_3$, ...) could not be observed in MQWs probably because of the larger inhomogeneity.

\begin{figure}[htb]
	\includegraphics[width=.4\textwidth]{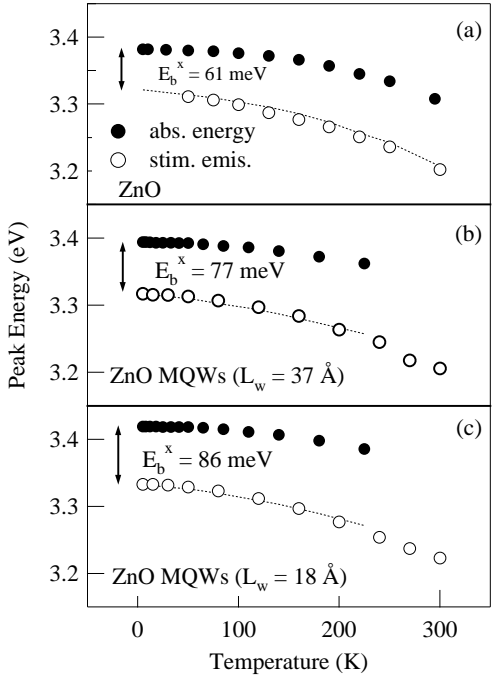}
\caption{Temperature dependences of peak energies of the \textit{P} bands (open
circles) and free exciton energies (closed circles) in a ZnO
epitaxial layer (a), ZnO/Zn$_{0.88}$Mg$_{0.12} $O (b) and ZnO/Zn$_{0.73}$Mg$_{0.27} $O MQWs (c)~\protect\cite{sun1}.}
\label{peak_pos}
\end{figure}

For an unambiguous identification, it is necessary to perform a systematic investigation. In order to facilitate the comparison of the experimental data with Eq.~(\ref{kinetic}) including the kinetic energy term, we plot in Figs.~\ref{peak_pos}(b) and (c) that correspond to the temperature dependence of the peak energy of the \textit{P} band in the QWs (Mg concentration of 0.12, $L_w$s of 37\textrm{\AA } and 18\textrm{\AA }). For comparison, a corresponding plot for a thin ZnO film is shown in Fig.~\ref{peak_pos}(a). At sufficiently low temperatures, the energy separation of the \textit{P} band from the resonance energy of exciton must be equal to the exciton binding energy because the kinetic energy approaches zero. The temperature dependence of the peak energy difference in QWs shows the same behavior as that of ZnO. Thus, as a result of such a careful comparison, it can be said that the mechanism of this stimulated emission for the MQWs is the inelastic x-x scattering. This also rules out the other possibilities such as stimulated exciton-phonon scattering as a mechanism leading to this excitonic gain. This method was also used for the determination of exciton binding energies. These estimated values have been already summarized in a previous section (cf. Fig.~\ref{bindingenergy} of section~\ref{sec:linear}), indicating the confinement-induced enhancement.

The x-x recombination process is not only observed in the luminescence, but also in the optical gain spectra. Sun \textit{et~al.} experimentally determined the value of the optical gain by using so-called the variable stripe-length method. The peak value of the excitonic gain of a ZnO/Zn$_{0.88}$Mg$_{0.12}$O MQW with $L_w = 18$~\textrm{\AA} at room temperature was estimated to be $\simeq$250~cm$^{-1}$.

According to results of the high-temperature tolerance tests, the stimulated emission could be observed up to 377~K, and the characteristic temperature, which is a figure of merit with respect to temperature rise, was estimated to be 87~K~\cite{ohtomo8}. This was significantly higher than that of a 55-nm-thick ZnO/sapphire (67~K)~\cite{ohtomophd}. The lowest threshold value at room temperature was 11~kW/cm$^{2}$ in the case of $L_w$ of 47~\textrm{\AA}. See Ref.~\cite{ohtomo8} for details.

\subsection{Optical properties of biexcitons}
\subsubsection{Radiative recombination from biexcitonic states}
\begin{figure}[htb]
	\includegraphics[width=.4\textwidth]{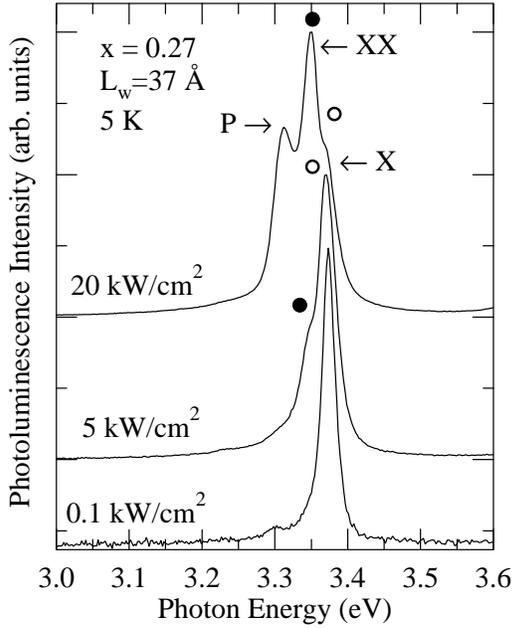}
		\caption{PL spectra at 5 K taken from a ZnO/ZnMgO MQW with well width of 18~\textrm{\AA} under three different excitation intensities.}
	\label{biexciton-pl}
\end{figure}

As has been introduced in a previous subsection, in principle the threshold for  laser action using biexcitons as its mechanism is expected to be even lower than that using an x-x scattering process. The reduced binding energy of biexciton with respect to the exciton however makes the biexciton particle rather unstable, so that the room-temperature achievement of biexciton lasing seems to be very difficult even in ZnO MQWs. Nevertheless, the formation of biexcitons has been already found at $T \le 160$~K in some ZnO QWs: the excitation-intensity dependence of the biexcitonic PL spectra and of nonlinear pump-probe spectra have been reported so far. Figure~\ref{biexciton-pl} gives such an example taken for the MQW (Mg concentration $x = 0.27$ and 37~\textrm{\AA}) at 5~K. We here would like to emphasize that the threshold for the biexciton recombination or formation is significantly lower than that for the x-x recombination, i.e., the biexcitonic gain is more advantageous from the view point of the application as a low-threshold semiconductor laser. The three traces in Fig.~\ref{biexciton-pl} show the normalized PL spectra at various excitation power under the optical excitation using a pulsed-dye laser (341~nm)~\cite{sun3}. 

At the lowest power density [100~W/cm$^2$, Fig.~\ref{biexciton-pl}], the luminescence spectrum is dominated by radiative recombination of localized excitons (an open circle, \textit{X}). As the excitation power density increases, there appears a shoulder (a closed circle, \textit{XX}) on the low-energy side of the \textit{X}-band. This emission band, located at 3.35~eV, grows superlinearly with respect to the excitation intensity. With further increase in excitation intensity, a second peak denoted by \textit{P} appears at around 3.12~eV. This assignment is supported by the rate-equation analyis for dependence of biexciton PL intensity on the excitation density (cf. Fig.~3 of Ref.~\cite{sun3}).

Probably because of the competition with the inelastic scattering processes, stimulated emission related to the biexciton decay has not been observed yet experimentally.

\subsubsection{Some other nonlinear spectroscopy and biexciton binding energy}

\begin{figure}[htb]
	\includegraphics[width=.4\textwidth]{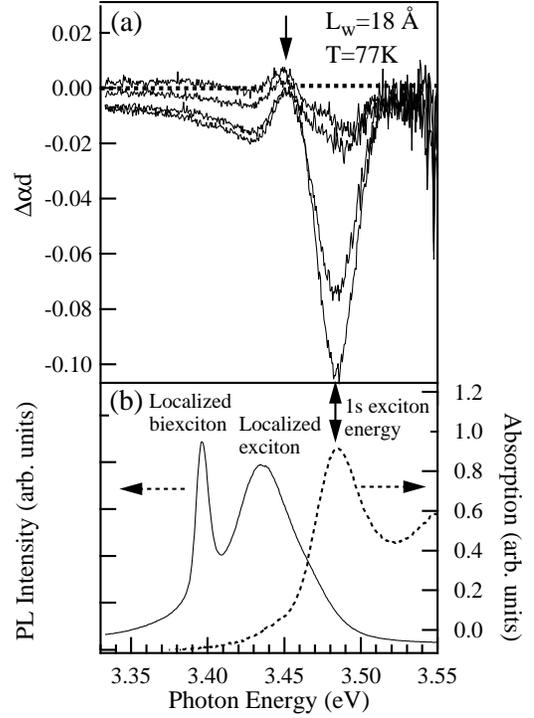}
	\caption{(a) Evolution of a differential absorption spectrum in a ZnO/Zn$_{0.73}$Mg$_{0.27}$O MQW ($L_w = 18$~\AA) with increasing excitation intensity taken at 77~K. (b) The PL and absorption spectra of the same MQW.}
	\label{differential}
\end{figure}

In this subsection we deal with the experimental results of a nonlinear spectroscopic method and the conclusions which may be drawn from these results concerning biexcitons. The method discussed here is two-step absorption into the biexciton states via really excited intermediate states of excitons.

\begin{figure}[htb]
	\includegraphics[width=.375\textwidth]{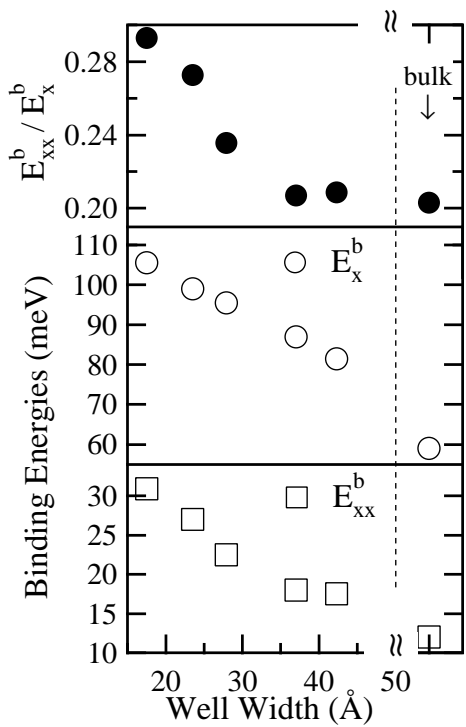}
	\caption{Well-width dependence of biexciton binding energy (open squares) estimated in ZnO/ZnMgO MQWs. Binding energies of free excitons (open circles), cited from Ref.~\cite{sunfull1}, and the $E^b_{BX} \slash E^b_{Ex}$ ratios (filled circles) as a function of the $L_w$, are also shown. We plotted the corresponding values of bulk ZnO for comparison at the right-hand side. The $E^b_{BX}$ and $E^b_{Ex}$ of bulk ZnO are, respectively, cited from Refs.~\cite{LBZincoxide} and \cite{hvam3}. Solid lines were drawn for visual guidance.}
	\label{biexbinding}
\end{figure}

In order to exploited the biexciton creation via the stepwise two-photon absorption and determine its binding energy, Chia \textit{et~al.} adopted a pump-probe transmission spectroscopy, in which a strong pump beam creates the high-density excitonic gas while a weak tunable beam probes the changes of the excitonic absorption. In their experiments, a ZnO MQW is illuminated by a XeCl excimer-laser beam, and a second broad-band light source prepared by a dye solution. The use of the laser beam exactly tuned to the resonance simplifies the situation and interpretation of the spectroscopic data. Under the condition of nonresonant excitation, densely distributed electron-hole pairs initially created are very rapidly relaxed to the exciton states, leading to the formation of high-density free excitons, the energy of which is equal to their transverse energy. A two photon transition or stimulated-absorption process manifests itself in a dip in the transmission spectrum, which satisfies the following relation:
\begin{equation}
\hbar \omega_{exc} + \hbar \omega_{dip} = E_m
\end{equation}
In addition to the bleaching phenomenon of absorption due to the saturation of excitonic states, they surely observed such induced absorption (cf. Fig~\ref{differential}). Moreover, they demonstrated that the transition involved in the induced absorption process is from free-excitonic states to free-biexcitonic states even in the case of ZnO quantum structures where the localization effect are not negligible. Therefore, this experimental method is found to be useful to precisely determine the binding energy of biexciton unlike the PL or stimulated emission spectroscopy. In the latter case as already stated in subsection~\ref{basictheoretical}, spectra are dominated by the radiative recombination of biexcitons localized at the potential variation caused by well width and depth fluctuations. No quantitative interpretation of these experiments has been performed yet. For small broadening, free biexciton binding energy can be estimated by using the energy separation between the exciton resonance and lower-energy induced absorption peak. By estimating the separation of these peak energies, the biexciton binding energies ($E^b_{BX}$) were plotted against $L_w$ (open squares) in Fig.~\ref{biexbinding}. For comparison, the dependence of exciton binding energy ($E^b_{Ex}$, open circles) on $L_w$, cited from Ref.~\cite{sunfull1}, is also shown. Both these binding energies are monotonically increasing functions of $L_w$. 

The ratios between these two binding energies are also shown by full circles for comparison. Lasing can occur through the stimulated recombination of biexcitons, provided an intermediate level exists to from the three level system, namely a biexciton state or the final state of a scattering process (an exciton state). Experimental results unambiguously evidencing the room-temperature biexcitonic gain are highly desirable by optimizing a QW design.

\subsection{Electron-hole plasma}

In this section we mention very briefly about the electron-hole plasmas which have been used as a recombination process for laser action of all the semiconductor lasers commercially available. The basic physics is analogous to the case of one-component plasmas that are distributed inside a modulation-doped semiconductors. Such an emission (usually called ``\textit{N}'' emission) has been observed in ZnO thin films and QWs at elevated pumping levels more intense than the case of x-x scattering. The QWs grown by a plasma-assisted MBE technique have exhibited only stimulated emission originating from the EHP phase.
The EHP recombination process is also observed in the optical gain spectra. The peak value of this optical gain was estimated to be $\simeq$390~cm$^{-1}$ at room temperature~\cite{sunspie2001}.

\section{Conclusions}
\label{sec:conclusion}
We have overviewed the main linear and nonlinear optical properties of excitons in ZnO-based QWs. The fundamental importance of strongly bound excitons in terms of localization of excitons, temporal dynamics, and oscillator strength has been addressed thought the discussion of recent spectroscopic experiments. We also reported on an influence of internal electric field present along the growth axis of the wurtzite semiconductor QWs, resulting in observation of PL of spatially separated electron-hole pairs due to the quantum-confined Stark effects in ZnO/Mg$_{0.27}$Zn$_{0.73}$O QWs. The role of excitons in the operation of optoelectronic devices operating in the blue ultraviolet regions has been emphasized in the discussion of the nonlinear optical properties. The unique combination of large exciton binding energy and reduced screening and phonon coupling occurring in QWs permits the observation of novel phenomena connected with the coexistence of the exciton gas and the free carrier gas at high density. This property is propitious to the stabilization of excitons at high temperatures. The confinement-induced enhancement of the exciton as well as biexciton binding energies were demonstrated in QWs. Besides their fundamental importance, these phenomena are found to play an important role in the physics of blue-ultraviolet lasers and modulators.

Such high-quality MQWs opens up numerous possibilities for UV optoelectric devices.

\section*{Acknowledgements}

The quantum wells used as the target of optical characterization in our study were produced by Dr.~A.~Ohtomo and K.~Tamura using equipment originally constructed by Dr.~Y.~Matsumoto, Tokyo Institute of Technology, Japan.
We would like to express our gratitude to the abovementioned researchers as well as Dr.~Ngyuen Tien Tuan, Dr.~H. D. Sun and C.~H. Chia, Institute of Physical and Chemical Research, Japan, for their assistance in all aspects of our research.


\begin{thebibliography}{10}
\expandafter\ifx\csname bibnamefont\endcsname\relax
  \def\bibnamefont#1{#1}\fi
\expandafter\ifx\csname bibfnamefont\endcsname\relax
  \def\bibfnamefont#1{#1}\fi
\expandafter\ifx\csname url\endcsname\relax
  \def\url#1{\texttt{#1}}\fi
\expandafter\ifx\csname urlprefix\endcsname\relax\def\urlprefix{URL }\fi
\providecommand{\bibinfo}[2]{#2}
\providecommand{\eprint}[2][]{\url{#2}}

\bibitem{makino8}
\bibinfo{author}{\bibfnamefont{T.}~\bibnamefont{Makino}},
  \bibinfo{author}{\bibfnamefont{C.~H.} \bibnamefont{Chia}},
  \bibinfo{author}{\bibfnamefont{N.~T.} \bibnamefont{Tuan}},
  \bibinfo{author}{\bibfnamefont{Y.}~\bibnamefont{Segawa}},
  \bibinfo{author}{\bibfnamefont{M.}~\bibnamefont{Kawasaki}},
  \bibinfo{author}{\bibfnamefont{A.}~\bibnamefont{Ohtomo}},
  \bibinfo{author}{\bibfnamefont{K.}~\bibnamefont{Tamura}}, \bibnamefont{and}
  \bibinfo{author}{\bibfnamefont{H.}~\bibnamefont{Koinuma}},
  \bibinfo{journal}{Appl. Phys. Lett.}
  \textbf{\bibinfo{volume}{76}}(\bibinfo{number}{24}), \bibinfo{pages}{3549}
  (\bibinfo{year}{2000}).

\bibitem{chen1}
\bibinfo{author}{\bibfnamefont{Y.~F.} \bibnamefont{Chen}},
  \bibinfo{author}{\bibfnamefont{H.~K.} \bibnamefont{Ko}},
  \bibinfo{author}{\bibfnamefont{S.}~\bibnamefont{Hong}}, \bibnamefont{and}
  \bibinfo{author}{\bibfnamefont{T.}~\bibnamefont{Yao}},
  \bibinfo{journal}{Appl. Phys. Lett.}
  \textbf{\bibinfo{volume}{76}}(\bibinfo{number}{5}), \bibinfo{pages}{559}
  (\bibinfo{year}{2000}).

\bibitem{gil2}
\bibinfo{author}{\bibfnamefont{B.}~\bibnamefont{Gil}}, \bibinfo{journal}{Phys.
  Rev. B} \textbf{\bibinfo{volume}{64}}(\bibinfo{number}{20}),
  \bibinfo{pages}{201310(R)} (\bibinfo{year}{2001}).

\bibitem{segawa1}
\bibinfo{author}{\bibfnamefont{Y.}~\bibnamefont{Segawa}},
  \bibinfo{author}{\bibfnamefont{A.}~\bibnamefont{Ohtomo}},
  \bibinfo{author}{\bibfnamefont{M.}~\bibnamefont{Kawasaki}},
  \bibinfo{author}{\bibfnamefont{H.}~\bibnamefont{Koinuma}},
  \bibinfo{author}{\bibfnamefont{Z.}~\bibnamefont{Tang}},
  \bibinfo{author}{\bibfnamefont{P.}~\bibnamefont{Yu}}, \bibnamefont{and}
  \bibinfo{author}{\bibfnamefont{G.}~\bibnamefont{Wong}},
  \bibinfo{journal}{Phys. Status Solidi (b)} \textbf{\bibinfo{volume}{202}},
  \bibinfo{pages}{609} (\bibinfo{year}{1997}).

\bibitem{nomura1}
\bibinfo{author}{\bibfnamefont{K.}~\bibnamefont{Nomura}},
  \bibinfo{author}{\bibfnamefont{H.}~\bibnamefont{Ohta}},
  \bibinfo{author}{\bibfnamefont{K.}~\bibnamefont{Ueda}},
  \bibinfo{author}{\bibfnamefont{T.}~\bibnamefont{Kamiya}},
  \bibinfo{author}{\bibfnamefont{M.}~\bibnamefont{Hirano}}, \bibnamefont{and}
  \bibinfo{author}{\bibfnamefont{H.}~\bibnamefont{Hosono}},
  \bibinfo{journal}{Science}
  \textbf{\bibinfo{volume}{300}}(\bibinfo{number}{5623}), \bibinfo{pages}{1269}
  (\bibinfo{year}{2003}).

\bibitem{carcia1}
\bibinfo{author}{\bibfnamefont{P.~F.} \bibnamefont{Carcia}},
  \bibinfo{author}{\bibfnamefont{R.~S.} \bibnamefont{Mclean}},
  \bibinfo{author}{\bibfnamefont{M.~H.} \bibnamefont{Reilly}},
  \bibnamefont{and} \bibinfo{author}{\bibfnamefont{G.}~\bibnamefont{Nunes}},
  \bibinfo{journal}{Appl. Phys. Lett.}
  \textbf{\bibinfo{volume}{82}}(\bibinfo{number}{7}), \bibinfo{pages}{1117}
  (\bibinfo{year}{2003}).

\bibitem{nishii1}
\bibinfo{author}{\bibfnamefont{J.}~\bibnamefont{Nishii}},
  \bibinfo{author}{\bibfnamefont{F.~M.} \bibnamefont{Hossain}},
  \bibinfo{author}{\bibfnamefont{S.}~\bibnamefont{Takagi}},
  \bibinfo{author}{\bibfnamefont{T.}~\bibnamefont{Aita}},
  \bibinfo{author}{\bibfnamefont{K.}~\bibnamefont{Saikusa}},
  \bibinfo{author}{\bibfnamefont{Y.}~\bibnamefont{Ohmaki}},
  \bibinfo{author}{\bibfnamefont{I.}~\bibnamefont{Ohkubo}},
  \bibinfo{author}{\bibfnamefont{S.}~\bibnamefont{Kishimoto}},
  \bibnamefont{and} \bibinfo{author}{\bibfnamefont{A.}~\bibnamefont{Ohtomo}},
  \bibinfo{journal}{Jpn. J. Appl. Phys.}
  \textbf{\bibinfo{volume}{42}}(\bibinfo{number}{4A}), \bibinfo{pages}{L347}
  (\bibinfo{year}{2003}).

\bibitem{masuda1}
\bibinfo{author}{\bibfnamefont{S.}~\bibnamefont{Masuda}},
  \bibinfo{author}{\bibfnamefont{K.}~\bibnamefont{Kitamura}},
  \bibinfo{author}{\bibfnamefont{Y.}~\bibnamefont{Okumura}},
  \bibinfo{author}{\bibfnamefont{S.}~\bibnamefont{Miyatake}},
  \bibinfo{author}{\bibfnamefont{H.}~\bibnamefont{Tabata}}, \bibnamefont{and}
  \bibinfo{author}{\bibfnamefont{T.}~\bibnamefont{Kawai}}, \bibinfo{journal}{J.
  Appl. Phys.} \textbf{\bibinfo{volume}{93}}(\bibinfo{number}{3}),
  \bibinfo{pages}{1624} (\bibinfo{year}{2003}).

\bibitem{hoffman1}
\bibinfo{author}{\bibfnamefont{R.~L.} \bibnamefont{Hoffman}},
  \bibinfo{author}{\bibfnamefont{B.~J.} \bibnamefont{Norris}},
  \bibnamefont{and} \bibinfo{author}{\bibfnamefont{J.~F.} \bibnamefont{Wager}},
  \bibinfo{journal}{Appl. Phys. Lett.}
  \textbf{\bibinfo{volume}{85}}(\bibinfo{number}{5}), \bibinfo{pages}{733}
  (\bibinfo{year}{2003}).

\bibitem{LBZincoxide}
\bibinfo{author}{\bibfnamefont{E.}~\bibnamefont{Mollwo}}, in
  \emph{\bibinfo{booktitle}{Semiconductors: Physics of II-VI and I-VII
  Compounds, Semimagnetic Semiconductors}}, edited by
  \bibinfo{editor}{\bibfnamefont{O.}~\bibnamefont{Madelung}},
  \bibinfo{editor}{\bibfnamefont{M.}~\bibnamefont{Schulz}}, \bibnamefont{and}
  \bibinfo{editor}{\bibfnamefont{H.}~\bibnamefont{Weiss}}
  (\bibinfo{publisher}{Springer}, \bibinfo{address}{Berlin},
  \bibinfo{year}{1982}), vol.~\bibinfo{volume}{17} of
  \emph{\bibinfo{series}{Landolt-B{\"{o}}rnstein New Series}},
  p.~\bibinfo{pages}{35}.

\bibitem{sun1}
\bibinfo{author}{\bibfnamefont{H.~D.} \bibnamefont{Sun}},
  \bibinfo{author}{\bibfnamefont{T.}~\bibnamefont{Makino}},
  \bibinfo{author}{\bibfnamefont{N.~T.} \bibnamefont{Tuan}},
  \bibinfo{author}{\bibfnamefont{Y.}~\bibnamefont{Segawa}},
  \bibinfo{author}{\bibfnamefont{Z.~K.} \bibnamefont{Tang}},
  \bibinfo{author}{\bibfnamefont{G.~K.~L.} \bibnamefont{Wong}},
  \bibinfo{author}{\bibfnamefont{M.}~\bibnamefont{Kawasaki}},
  \bibinfo{author}{\bibfnamefont{A.}~\bibnamefont{Ohtomo}},
  \bibinfo{author}{\bibfnamefont{K.}~\bibnamefont{Tamura}}, \bibnamefont{and}
  \bibinfo{author}{\bibfnamefont{H.}~\bibnamefont{Koinuma}},
  \bibinfo{journal}{Appl. Phys. Lett.}
  \textbf{\bibinfo{volume}{77}}(\bibinfo{number}{26}), \bibinfo{pages}{4250}
  (\bibinfo{year}{2000}).

\bibitem{sunfull1}
\bibinfo{author}{\bibfnamefont{H.~D.} \bibnamefont{Sun}},
  \bibinfo{author}{\bibfnamefont{T.}~\bibnamefont{Makino}},
  \bibinfo{author}{\bibfnamefont{Y.}~\bibnamefont{Segawa}},
  \bibinfo{author}{\bibfnamefont{M.}~\bibnamefont{Kawasaki}},
  \bibinfo{author}{\bibfnamefont{A.}~\bibnamefont{Ohtomo}},
  \bibinfo{author}{\bibfnamefont{K.}~\bibnamefont{Tamura}}, \bibnamefont{and}
  \bibinfo{author}{\bibfnamefont{H.}~\bibnamefont{Koinuma}},
  \bibinfo{journal}{J. Appl. Phys.}
  \textbf{\bibinfo{volume}{91}}(\bibinfo{number}{4}), \bibinfo{pages}{1993}
  (\bibinfo{year}{2002}).

\bibitem{sun2}
\bibinfo{author}{\bibfnamefont{H.~D.} \bibnamefont{Sun}},
  \bibinfo{author}{\bibfnamefont{T.}~\bibnamefont{Makino}},
  \bibinfo{author}{\bibfnamefont{N.~T.} \bibnamefont{Tuan}},
  \bibinfo{author}{\bibfnamefont{Y.}~\bibnamefont{Segawa}},
  \bibinfo{author}{\bibfnamefont{M.}~\bibnamefont{Kawasaki}},
  \bibinfo{author}{\bibfnamefont{A.}~\bibnamefont{Ohtomo}},
  \bibinfo{author}{\bibfnamefont{K.}~\bibnamefont{Tamura}}, \bibnamefont{and}
  \bibinfo{author}{\bibfnamefont{H.}~\bibnamefont{Koinuma}},
  \bibinfo{journal}{Appl. Phys. Lett.}
  \textbf{\bibinfo{volume}{78}}(\bibinfo{number}{17}), \bibinfo{pages}{2464}
  (\bibinfo{year}{2001}).

\bibitem{makino11}
\bibinfo{author}{\bibfnamefont{T.}~\bibnamefont{Makino}},
  \bibinfo{author}{\bibfnamefont{N.~T.} \bibnamefont{Tuan}},
  \bibinfo{author}{\bibfnamefont{H.~D.} \bibnamefont{Sun}},
  \bibinfo{author}{\bibfnamefont{C.~H.} \bibnamefont{Chia}},
  \bibinfo{author}{\bibfnamefont{Y.}~\bibnamefont{Segawa}},
  \bibinfo{author}{\bibfnamefont{M.}~\bibnamefont{Kawasaki}},
  \bibinfo{author}{\bibfnamefont{A.}~\bibnamefont{Ohtomo}},
  \bibinfo{author}{\bibfnamefont{K.}~\bibnamefont{Tamura}}, \bibnamefont{and}
  \bibinfo{author}{\bibfnamefont{H.}~\bibnamefont{Koinuma}},
  \bibinfo{journal}{Appl. Phys. Lett.}
  \textbf{\bibinfo{volume}{77}}(\bibinfo{number}{7}), \bibinfo{pages}{975}
  (\bibinfo{year}{2000}).

\bibitem{bernardini1}
\bibinfo{author}{\bibfnamefont{F.}~\bibnamefont{Bernardini}},
  \bibinfo{author}{\bibfnamefont{V.}~\bibnamefont{Fiorentini}},
  \bibnamefont{and}
  \bibinfo{author}{\bibfnamefont{D.}~\bibnamefont{Vanderbilt}},
  \bibinfo{journal}{Phys. Rev. Lett.}
  \textbf{\bibinfo{volume}{79}}(\bibinfo{number}{20}), \bibinfo{pages}{3958}
  (\bibinfo{year}{1997}).

\bibitem{seoung-hwanpark}
\bibinfo{author}{\bibfnamefont{S.-H.} \bibnamefont{Park}} \bibnamefont{and}
  \bibinfo{author}{\bibfnamefont{S.-L.} \bibnamefont{Chuang}},
  \bibinfo{journal}{J. Appl. Phys.}
  \textbf{\bibinfo{volume}{87}}(\bibinfo{number}{1}), \bibinfo{pages}{353}
  (\bibinfo{year}{2000}).

\bibitem{cingolani2}
\bibinfo{author}{\bibfnamefont{R.}~\bibnamefont{Cingolani}},
  \bibinfo{author}{\bibfnamefont{G.}~\bibnamefont{Larocca}},
  \bibinfo{author}{\bibfnamefont{H.}~\bibnamefont{Kalt}},
  \bibinfo{author}{\bibfnamefont{K.}~\bibnamefont{Ploog}}, \bibnamefont{and}
  \bibinfo{author}{\bibfnamefont{J.}~\bibnamefont{Mann}},
  \bibinfo{journal}{Phys. Rev. B} \textbf{\bibinfo{volume}{43}},
  \bibinfo{pages}{9662} (\bibinfo{year}{1991}).

\bibitem{langer1}
\bibinfo{author}{\bibfnamefont{R.}~\bibnamefont{Langer}},
  \bibinfo{author}{\bibfnamefont{J.}~\bibnamefont{Simon}},
  \bibinfo{author}{\bibfnamefont{V.}~\bibnamefont{Ortiz}},
  \bibinfo{author}{\bibfnamefont{N.~T.} \bibnamefont{Pelekanos}}, ,
  \bibinfo{author}{\bibfnamefont{A.}~\bibnamefont{Barski}},
  \bibinfo{author}{\bibfnamefont{R.}~\bibnamefont{Andre}}, \bibnamefont{and}
  \bibinfo{author}{\bibfnamefont{M.}~\bibnamefont{Godlewski}},
  \bibinfo{journal}{Appl. Phys. Lett.}
  \textbf{\bibinfo{volume}{74}}(\bibinfo{number}{25}), \bibinfo{pages}{3827}
  (\bibinfo{year}{1999}).

\bibitem{m_leroux1}
\bibinfo{author}{\bibfnamefont{M.}~\bibnamefont{Leroux}},
  \bibinfo{author}{\bibfnamefont{N.}~\bibnamefont{Grandjean}},
  \bibinfo{author}{\bibfnamefont{M.}~\bibnamefont{Laugt}},
  \bibinfo{author}{\bibfnamefont{J.}~\bibnamefont{Massies}},
  \bibinfo{author}{\bibfnamefont{B.}~\bibnamefont{Gil}},
  \bibinfo{author}{\bibfnamefont{P.}~\bibnamefont{Lefebvre}}, \bibnamefont{and}
  \bibinfo{author}{\bibfnamefont{P.}~\bibnamefont{Bigenwald}},
  \bibinfo{journal}{Phys. Rev. B}
  \textbf{\bibinfo{volume}{58}}(\bibinfo{number}{20}), \bibinfo{pages}{R13371}
  (\bibinfo{year}{1998}).

\bibitem{krishnamoorthy}
\bibinfo{author}{\bibfnamefont{S.}~\bibnamefont{Krishnamorthy}},
  \bibinfo{author}{\bibfnamefont{A.~A.} \bibnamefont{Iliadis}},
  \bibinfo{author}{\bibfnamefont{A.}~\bibnamefont{Inumpudi}},
  \bibinfo{author}{\bibfnamefont{S.}~\bibnamefont{Choopun}},
  \bibinfo{author}{\bibfnamefont{R.~D.} \bibnamefont{Vispute}},
  \bibnamefont{and}
  \bibinfo{author}{\bibfnamefont{T.}~\bibnamefont{Venkatesan}},
  \bibinfo{journal}{Solid State Electron.}
  \textbf{\bibinfo{volume}{46}}(\bibinfo{number}{10}), \bibinfo{pages}{1633}
  (\bibinfo{year}{2002}).

\bibitem{makino14}
\bibinfo{author}{\bibfnamefont{T.}~\bibnamefont{Makino}},
  \bibinfo{author}{\bibfnamefont{Y.}~\bibnamefont{Segawa}},
  \bibinfo{author}{\bibfnamefont{M.}~\bibnamefont{Kawasaki}},
  \bibinfo{author}{\bibfnamefont{A.}~\bibnamefont{Ohtomo}},
  \bibinfo{author}{\bibfnamefont{R.}~\bibnamefont{Shiroki}},
  \bibinfo{author}{\bibfnamefont{K.}~\bibnamefont{Tamura}},
  \bibinfo{author}{\bibfnamefont{T.}~\bibnamefont{Yasuda}}, \bibnamefont{and}
  \bibinfo{author}{\bibfnamefont{H.}~\bibnamefont{Koinuma}},
  \bibinfo{journal}{Appl. Phys. Lett.}
  \textbf{\bibinfo{volume}{78}}(\bibinfo{number}{9}), \bibinfo{pages}{1237}
  (\bibinfo{year}{2001}).

\bibitem{kawasaki2}
\bibinfo{author}{\bibfnamefont{M.}~\bibnamefont{Kawasaki}},
  \bibinfo{author}{\bibfnamefont{A.}~\bibnamefont{Ohtomo}},
  \bibinfo{author}{\bibfnamefont{R.}~\bibnamefont{Shiroki}},
  \bibinfo{author}{\bibfnamefont{I.}~\bibnamefont{Ohkubo}},
  \bibinfo{author}{\bibfnamefont{H.}~\bibnamefont{Kimura}},
  \bibinfo{author}{\bibfnamefont{G.}~\bibnamefont{Isoya}},
  \bibinfo{author}{\bibfnamefont{T.}~\bibnamefont{Yasuda}},
  \bibinfo{author}{\bibfnamefont{Y.}~\bibnamefont{Segawa}}, \bibnamefont{and}
  \bibinfo{author}{\bibfnamefont{H.}~\bibnamefont{Koinuma}}, in
  \emph{\bibinfo{booktitle}{Extended Abstracts of the 1998 International
  Conference on Solid State Devices and Materials}}
  (\bibinfo{publisher}{Business Ctr. Acad. Soc. Jpn.},
  \bibinfo{address}{Hiroshima, Japan}, \bibinfo{year}{1998}), p.
  \bibinfo{pages}{356}.

\bibitem{vispute1}
\bibinfo{author}{\bibfnamefont{R.~D.} \bibnamefont{Vispute}},
  \bibinfo{author}{\bibfnamefont{V.}~\bibnamefont{Talyansky}},
  \bibinfo{author}{\bibfnamefont{S.}~\bibnamefont{Choopun}},
  \bibinfo{author}{\bibfnamefont{R.~P.} \bibnamefont{Sharma}},
  \bibinfo{author}{\bibfnamefont{T.}~\bibnamefont{Venkatesan}},
  \bibinfo{author}{\bibfnamefont{M.}~\bibnamefont{He}},
  \bibinfo{author}{\bibfnamefont{X.}~\bibnamefont{Tang}},
  \bibinfo{author}{\bibfnamefont{J.~B.} \bibnamefont{Halpern}},
  \bibinfo{author}{\bibfnamefont{M.~G.} \bibnamefont{Spencer}},
  \bibinfo{author}{\bibfnamefont{Y.~X.} \bibnamefont{Li}},
  \bibinfo{author}{\bibfnamefont{L.~G.} \bibnamefont{Salamanca-Riba}},
  \bibinfo{author}{\bibfnamefont{A.~A.} \bibnamefont{Iliadis}}, \emph{et~al.},
  \bibinfo{journal}{Appl. Phys. Lett.}
  \textbf{\bibinfo{volume}{73}}(\bibinfo{number}{3}), \bibinfo{pages}{348}
  (\bibinfo{year}{1998}).

\bibitem{narayan1}
\bibinfo{author}{\bibfnamefont{J.}~\bibnamefont{Narayan}},
  \bibinfo{author}{\bibfnamefont{K.}~\bibnamefont{Dovidenko}},
  \bibinfo{author}{\bibfnamefont{A.~K.} \bibnamefont{Sharma}},
  \bibnamefont{and}
  \bibinfo{author}{\bibfnamefont{S.}~\bibnamefont{Oktyabrsky}},
  \bibinfo{journal}{J. Appl. Phys.}
  \textbf{\bibinfo{volume}{84}}(\bibinfo{number}{5}), \bibinfo{pages}{2597}
  (\bibinfo{year}{1998}).

\bibitem{ohkubo1}
\bibinfo{author}{\bibfnamefont{I.}~\bibnamefont{Ohkubo}},
  \bibinfo{author}{\bibfnamefont{C.}~\bibnamefont{Hirose}},
  \bibinfo{author}{\bibfnamefont{K.}~\bibnamefont{Tamura}},
  \bibinfo{author}{\bibfnamefont{J.}~\bibnamefont{Nishii}},
  \bibinfo{author}{\bibfnamefont{H.}~\bibnamefont{Saito}},
  \bibinfo{author}{\bibfnamefont{H.}~\bibnamefont{Koinuma}},
  \bibinfo{author}{\bibfnamefont{P.}~\bibnamefont{Ahmet}},
  \bibinfo{author}{\bibfnamefont{T.}~\bibnamefont{Chikyouw}},
  \bibinfo{author}{\bibfnamefont{T.}~\bibnamefont{Ishii}},
  \bibinfo{author}{\bibfnamefont{S.}~\bibnamefont{Miyazawa}},
  \bibinfo{author}{\bibfnamefont{Y.}~\bibnamefont{Segawa}},
  \bibinfo{author}{\bibfnamefont{T.}~\bibnamefont{Fukumura}}, \emph{et~al.},
  \bibinfo{journal}{J. Appl. Phys.}
  \textbf{\bibinfo{volume}{92}}(\bibinfo{number}{0}), \bibinfo{pages}{5587}
  (\bibinfo{year}{2002}).

\bibitem{edahiro1}
\bibinfo{author}{\bibfnamefont{T.}~\bibnamefont{Edahiro}},
  \bibinfo{author}{\bibfnamefont{N.}~\bibnamefont{Fujimura}}, \bibnamefont{and}
  \bibinfo{author}{\bibfnamefont{T.}~\bibnamefont{Ito}}, \bibinfo{journal}{J.
  Appl. Phys.} \textbf{\bibinfo{volume}{93}}(\bibinfo{number}{10}),
  \bibinfo{pages}{7673} (\bibinfo{year}{2003}).

\bibitem{bogatu1}
\bibinfo{author}{\bibfnamefont{V.}~\bibnamefont{Bogatu}},
  \bibinfo{author}{\bibfnamefont{A.}~\bibnamefont{Goldenblum}},
  \bibinfo{author}{\bibfnamefont{A.}~\bibnamefont{Many}}, \bibnamefont{and}
  \bibinfo{author}{\bibfnamefont{Y.}~\bibnamefont{Goldstein}},
  \bibinfo{journal}{Phys. Status Solidi (b)}
  \textbf{\bibinfo{volume}{212}}(\bibinfo{number}{1}), \bibinfo{pages}{89}
  (\bibinfo{year}{1999}).

\bibitem{chen-jvb}
\bibinfo{author}{\bibfnamefont{Y.}~\bibnamefont{Chen}},
  \bibinfo{author}{\bibfnamefont{H.}~\bibnamefont{Ko}},
  \bibinfo{author}{\bibfnamefont{S.}~\bibnamefont{Hong}},
  \bibinfo{author}{\bibfnamefont{T.}~\bibnamefont{Sekiuchi}},
  \bibinfo{author}{\bibfnamefont{T.}~\bibnamefont{Yao}}, \bibnamefont{and}
  \bibinfo{author}{\bibfnamefont{Y.}~\bibnamefont{Segawa}},
  \bibinfo{journal}{J. Vac. Sci. Technol. B}
  \textbf{\bibinfo{volume}{18}}(\bibinfo{number}{3}), \bibinfo{pages}{1514}
  (\bibinfo{year}{2000}).

\bibitem{matsumoto1}
\bibinfo{author}{\bibfnamefont{Y.}~\bibnamefont{Matsumoto}},
  \bibinfo{author}{\bibfnamefont{M.}~\bibnamefont{Murakami}},
  \bibinfo{author}{\bibfnamefont{Z.~W.} \bibnamefont{Jin}},
  \bibinfo{author}{\bibfnamefont{A.}~\bibnamefont{Ohtomo}},
  \bibinfo{author}{\bibfnamefont{M.}~\bibnamefont{Lippmaa}},
  \bibinfo{author}{\bibfnamefont{M.}~\bibnamefont{Kawasaki}}, \bibnamefont{and}
  \bibinfo{author}{\bibfnamefont{H.}~\bibnamefont{Koinuma}},
  \bibinfo{journal}{Jpn. J. Appl. Phys., Part2}
  \textbf{\bibinfo{volume}{38}}(\bibinfo{number}{2(6A/B)}),
  \bibinfo{pages}{L603} (\bibinfo{year}{1999}).

\bibitem{coli1}
\bibinfo{author}{\bibfnamefont{G.}~\bibnamefont{Coli}} \bibnamefont{and}
  \bibinfo{author}{\bibfnamefont{K.~K.} \bibnamefont{Bajaj}},
  \bibinfo{journal}{Appl. Phys. Lett.}
  \textbf{\bibinfo{volume}{78}}(\bibinfo{number}{19}), \bibinfo{pages}{2861}
  (\bibinfo{year}{2001}).

\bibitem{ohtomo4}
\bibinfo{author}{\bibfnamefont{A.}~\bibnamefont{Ohtomo}},
  \bibinfo{author}{\bibfnamefont{M.}~\bibnamefont{Kawasaki}},
  \bibinfo{author}{\bibfnamefont{I.}~\bibnamefont{Ohkubo}},
  \bibinfo{author}{\bibfnamefont{H.}~\bibnamefont{Koinuma}},
  \bibinfo{author}{\bibfnamefont{T.}~\bibnamefont{Yasuda}}, \bibnamefont{and}
  \bibinfo{author}{\bibfnamefont{Y.}~\bibnamefont{Segawa}},
  \bibinfo{journal}{Appl. Phys. Lett.}
  \textbf{\bibinfo{volume}{75}}(\bibinfo{number}{1}), \bibinfo{pages}{980}
  (\bibinfo{year}{1999}).

\bibitem{pollmann1}
\bibinfo{author}{\bibfnamefont{J.}~\bibnamefont{Pollmann}} \bibnamefont{and}
  \bibinfo{author}{\bibfnamefont{H.}~\bibnamefont{B{\"u}ttner}},
  \bibinfo{journal}{Phys. Rev. B}
  \textbf{\bibinfo{volume}{16}}(\bibinfo{number}{10}), \bibinfo{pages}{4480}
  (\bibinfo{year}{1977}).

\bibitem{schwarz1}
\bibinfo{author}{\bibfnamefont{R.~B.} \bibnamefont{Schwarz}},
  \bibinfo{author}{\bibfnamefont{K.}~\bibnamefont{Khachaturyan}},
  \bibnamefont{and} \bibinfo{author}{\bibfnamefont{E.~R.} \bibnamefont{Weber}},
  \bibinfo{journal}{Appl. Phys. Lett.}
  \textbf{\bibinfo{volume}{70}}(\bibinfo{number}{9}), \bibinfo{pages}{1122}
  (\bibinfo{year}{1997}).

\bibitem{goergens1}
\bibinfo{author}{\bibfnamefont{L.}~\bibnamefont{G{\"o}rgens}},
  \bibinfo{author}{\bibfnamefont{O.}~\bibnamefont{Ambacher}},
  \bibinfo{author}{\bibfnamefont{M.}~\bibnamefont{Stutzmann}},
  \bibinfo{author}{\bibfnamefont{C.}~\bibnamefont{Miskys}},
  \bibinfo{author}{\bibfnamefont{F.}~\bibnamefont{Scholz}}, \bibnamefont{and}
  \bibinfo{author}{\bibfnamefont{J.}~\bibnamefont{Off}},
  \bibinfo{journal}{Appl. Phys. Lett.}
  \textbf{\bibinfo{volume}{76}}(\bibinfo{number}{5}), \bibinfo{pages}{577}
  (\bibinfo{year}{2000}).

\bibitem{bykhovski1}
\bibinfo{author}{\bibfnamefont{A.}~\bibnamefont{Bykhovski}},
  \bibinfo{author}{\bibfnamefont{B.}~\bibnamefont{Gelmont}}, \bibnamefont{and}
  \bibinfo{author}{\bibfnamefont{M.}~\bibnamefont{Shur}}, \bibinfo{journal}{J.
  Appl. Phys.} \textbf{\bibinfo{volume}{74}}(\bibinfo{number}{(11)}),
  \bibinfo{pages}{6734} (\bibinfo{year}{1993}).

\bibitem{makino7}
\bibinfo{author}{\bibfnamefont{T.}~\bibnamefont{Makino}},
  \bibinfo{author}{\bibfnamefont{G.}~\bibnamefont{Isoya}},
  \bibinfo{author}{\bibfnamefont{Y.}~\bibnamefont{Segawa}},
  \bibinfo{author}{\bibfnamefont{C.~H.} \bibnamefont{Chia}},
  \bibinfo{author}{\bibfnamefont{T.}~\bibnamefont{Yasuda}},
  \bibinfo{author}{\bibfnamefont{M.}~\bibnamefont{Kawasaki}},
  \bibinfo{author}{\bibfnamefont{A.}~\bibnamefont{Ohtomo}},
  \bibinfo{author}{\bibfnamefont{K.}~\bibnamefont{Tamura}}, \bibnamefont{and}
  \bibinfo{author}{\bibfnamefont{H.}~\bibnamefont{Koinuma}},
  \bibinfo{journal}{J. Cryst. Growth} \textbf{\bibinfo{volume}{214/215}},
  \bibinfo{pages}{289} (\bibinfo{year}{2000}).

\bibitem{ohtomo7}
\bibinfo{author}{\bibfnamefont{A.}~\bibnamefont{Ohtomo}},
  \bibinfo{author}{\bibfnamefont{R.}~\bibnamefont{Shiroki}},
  \bibinfo{author}{\bibfnamefont{I.}~\bibnamefont{Ohkubo}},
  \bibinfo{author}{\bibfnamefont{H.}~\bibnamefont{Koinuma}}, \bibnamefont{and}
  \bibinfo{author}{\bibfnamefont{M.}~\bibnamefont{Kawasaki}},
  \bibinfo{journal}{Appl. Phys. Lett.}
  \textbf{\bibinfo{volume}{75}}(\bibinfo{number}{26}), \bibinfo{pages}{4088}
  (\bibinfo{year}{1999}).

\bibitem{cingolanirev}
\bibinfo{author}{\bibfnamefont{R.}~\bibnamefont{Cingolani}},
  \emph{\bibinfo{title}{Optical properties of excitons in ZnSe based quantum
  well heterostructures}} (\bibinfo{publisher}{Academic Press INC, SAN DIEGO},
  \bibinfo{year}{1997}), vol.~\bibinfo{volume}{44}, chap.~\bibinfo{chapter}{3},
  p. \bibinfo{pages}{163}.

\bibitem{dukeandmahan}
\bibinfo{author}{\bibfnamefont{C.~B.} \bibnamefont{Duke}} \bibnamefont{and}
  \bibinfo{author}{\bibfnamefont{G.~D.} \bibnamefont{Mahan}},
  \bibinfo{journal}{Phys. Rev.}
  \textbf{\bibinfo{volume}{139}}(\bibinfo{number}{6A}), \bibinfo{pages}{A1965}
  (\bibinfo{year}{1965}).

\bibitem{rudinandsegall1}
\bibinfo{author}{\bibfnamefont{S.}~\bibnamefont{Rudin}},
  \bibinfo{author}{\bibfnamefont{T.~L.} \bibnamefont{Reinecke}},
  \bibnamefont{and} \bibinfo{author}{\bibfnamefont{B.}~\bibnamefont{Segall}},
  \bibinfo{journal}{Phys. Rev. B}
  \textbf{\bibinfo{volume}{42}}(\bibinfo{number}{17}), \bibinfo{pages}{11218}
  (\bibinfo{year}{1990}).

\bibitem{pelekanos1}
\bibinfo{author}{\bibfnamefont{N.~T.} \bibnamefont{Pelekanos}},
  \bibinfo{author}{\bibfnamefont{J.}~\bibnamefont{Ding}},
  \bibinfo{author}{\bibfnamefont{M.}~\bibnamefont{Hagerott}},
  \bibinfo{author}{\bibfnamefont{A.~V.} \bibnamefont{Nurmikko}},
  \bibinfo{author}{\bibfnamefont{H.}~\bibnamefont{Luo}},
  \bibinfo{author}{\bibfnamefont{N.}~\bibnamefont{Samarth}}, \bibnamefont{and}
  \bibinfo{author}{\bibfnamefont{J.~K.} \bibnamefont{Furdyna}},
  \bibinfo{journal}{Phys. Rev. B} \textbf{\bibinfo{volume}{45}},
  \bibinfo{pages}{6037} (\bibinfo{year}{1992}).

\bibitem{alperovich1}
\bibinfo{author}{\bibfnamefont{V.~L.} \bibnamefont{Alperovich}},
  \bibinfo{author}{\bibfnamefont{V.~M.} \bibnamefont{Zaletin}},
  \bibinfo{author}{\bibnamefont{a.~f. kravchenko}}, \bibnamefont{and}
  \bibinfo{author}{\bibfnamefont{A.~S.} \bibnamefont{Terekhov}},
  \bibinfo{journal}{Phys. Status Solidi (B)}
  \textbf{\bibinfo{volume}{77}}(\bibinfo{number}{2}), \bibinfo{pages}{465}
  (\bibinfo{year}{1976}).

\bibitem{fischer2}
\bibinfo{author}{\bibfnamefont{A.~J.} \bibnamefont{Fischer}},
  \bibinfo{author}{\bibfnamefont{D.~S.} \bibnamefont{Kim}},
  \bibinfo{author}{\bibfnamefont{J.}~\bibnamefont{Hays}},
  \bibinfo{author}{\bibfnamefont{W.}~\bibnamefont{Shan}},
  \bibinfo{author}{\bibfnamefont{J.~J.} \bibnamefont{Song}},
  \bibinfo{author}{\bibfnamefont{D.~B.} \bibnamefont{Eason}},
  \bibinfo{author}{\bibfnamefont{J.}~\bibnamefont{Ren}},
  \bibinfo{author}{\bibfnamefont{J.~F.} \bibnamefont{Schetzina}},
  \bibinfo{author}{\bibfnamefont{H.}~\bibnamefont{Luo}},
  \bibinfo{author}{\bibfnamefont{J.~K.} \bibnamefont{Furdyna}},
  \bibinfo{author}{\bibfnamefont{Z.~Q.} \bibnamefont{Zhu}},
  \bibinfo{author}{\bibfnamefont{T.}~\bibnamefont{Yao}}, \emph{et~al.},
  \bibinfo{journal}{Phys. Rev. Lett.}
  \textbf{\bibinfo{volume}{73}}(\bibinfo{number}{17}), \bibinfo{pages}{2368}
  (\bibinfo{year}{1994}).

\bibitem{LeeJohnsonAustin}
\bibinfo{author}{\bibfnamefont{D.}~\bibnamefont{Lee}},
  \bibinfo{author}{\bibfnamefont{A.~M.} \bibnamefont{Johnson}},
  \bibinfo{author}{\bibfnamefont{J.~E.} \bibnamefont{Zucker}},
  \bibinfo{author}{\bibfnamefont{R.~D.} \bibnamefont{Feldman}},
  \bibnamefont{and} \bibinfo{author}{\bibfnamefont{R.~F.}
  \bibnamefont{Austin}}, \bibinfo{journal}{J. Appl. Phys.}
  \textbf{\bibinfo{volume}{69}}(\bibinfo{number}{9}), \bibinfo{pages}{6722}
  (\bibinfo{year}{1991}).

\bibitem{xbzhangGaNrev}
\bibinfo{author}{\bibfnamefont{X.~B.} \bibnamefont{Zhang}},
  \bibinfo{author}{\bibfnamefont{T.}~\bibnamefont{Taliercio}},
  \bibinfo{author}{\bibfnamefont{S.}~\bibnamefont{Kolliakos}},
  \bibnamefont{and} \bibinfo{author}{\bibfnamefont{P.}~\bibnamefont{Lefebvre}},
  \bibinfo{journal}{J. Phys.: Condens. Matter}
  \textbf{\bibinfo{volume}{13}}(\bibinfo{number}{1}), \bibinfo{pages}{7053}
  (\bibinfo{year}{2001}).

\bibitem{goni1}
\bibinfo{author}{\bibfnamefont{A.~R.} \bibnamefont{Goni}},
  \bibinfo{author}{\bibfnamefont{A.}~\bibnamefont{Cantarero}},
  \bibinfo{author}{\bibfnamefont{K.}~\bibnamefont{Syassen}}, \bibnamefont{and}
  \bibinfo{author}{\bibfnamefont{M.}~\bibnamefont{Cardona}},
  \bibinfo{journal}{Phys. Rev. B}
  \textbf{\bibinfo{volume}{41}}(\bibinfo{number}{14}), \bibinfo{pages}{10111}
  (\bibinfo{year}{1990}).

\bibitem{gurioli1}
\bibinfo{author}{\bibfnamefont{M.}~\bibnamefont{Gurioli}},
  \bibinfo{author}{\bibfnamefont{J.}~\bibnamefont{MartinezPastor}},
  \bibinfo{author}{\bibfnamefont{M.}~\bibnamefont{Colocci}},
  \bibinfo{author}{\bibfnamefont{A.}~\bibnamefont{Bosacchi}},
  \bibinfo{author}{\bibfnamefont{S.}~\bibnamefont{Franchi}}, \bibnamefont{and}
  \bibinfo{author}{\bibfnamefont{L.~C.} \bibnamefont{Andreani}},
  \bibinfo{journal}{Phys. Rev. B}
  \textbf{\bibinfo{volume}{47}}(\bibinfo{number}{23}), \bibinfo{pages}{15755}
  (\bibinfo{year}{1993}).

\bibitem{shinada1}
\bibinfo{author}{\bibfnamefont{M.}~\bibnamefont{Shinada}} \bibnamefont{and}
  \bibinfo{author}{\bibfnamefont{S.}~\bibnamefont{Sagano}},
  \bibinfo{journal}{J. Phys. Soc. Jpn.}
  \textbf{\bibinfo{volume}{21}}(\bibinfo{number}{10}), \bibinfo{pages}{1936}
  (\bibinfo{year}{1966}).

\bibitem{gourdon1}
\bibinfo{author}{\bibfnamefont{C.}~\bibnamefont{Gourdon}} \bibnamefont{and}
  \bibinfo{author}{\bibfnamefont{P.}~\bibnamefont{Lavallard}},
  \bibinfo{journal}{Phys. Status Solidi (b)} \textbf{\bibinfo{volume}{153}},
  \bibinfo{pages}{641} (\bibinfo{year}{1989}).

\bibitem{chia2}
\bibinfo{author}{\bibfnamefont{C.~H.} \bibnamefont{Chia}},
  \bibinfo{author}{\bibfnamefont{T.}~\bibnamefont{Makino}},
  \bibinfo{author}{\bibfnamefont{Y.}~\bibnamefont{Segawa}},
  \bibinfo{author}{\bibfnamefont{M.}~\bibnamefont{Kawasaki}},
  \bibinfo{author}{\bibfnamefont{A.}~\bibnamefont{Ohtomo}},
  \bibinfo{author}{\bibfnamefont{K.}~\bibnamefont{Tamura}}, \bibnamefont{and}
  \bibinfo{author}{\bibfnamefont{H.}~\bibnamefont{Koinuma}},
  \bibinfo{journal}{J. Appl. Phys.}
  \textbf{\bibinfo{volume}{90}}(\bibinfo{number}{7}), \bibinfo{pages}{3650}
  (\bibinfo{year}{2001}).

\bibitem{segallandmahan}
\bibinfo{author}{\bibfnamefont{B.}~\bibnamefont{Segall}} \bibnamefont{and}
  \bibinfo{author}{\bibfnamefont{G.~D.} \bibnamefont{Mahan}},
  \bibinfo{journal}{Phys. Rev.} \textbf{\bibinfo{volume}{171}},
  \bibinfo{pages}{935} (\bibinfo{year}{1968}).

\bibitem{gross2}
\bibinfo{author}{\bibfnamefont{E.}~\bibnamefont{Gross}},
  \bibinfo{author}{\bibfnamefont{S.}~\bibnamefont{Permogorov}},
  \bibnamefont{and} \bibinfo{author}{\bibfnamefont{B.}~\bibnamefont{Razbirin}},
  \bibinfo{journal}{J. Phys. Chem. Solids} \textbf{\bibinfo{volume}{27}},
  \bibinfo{pages}{1647} (\bibinfo{year}{1966}).

\bibitem{weiher1}
\bibinfo{author}{\bibfnamefont{R.~L.} \bibnamefont{Weiher}} \bibnamefont{and}
  \bibinfo{author}{\bibfnamefont{W.~C.} \bibnamefont{Tait}},
  \bibinfo{journal}{Phys. Rev.}
  \textbf{\bibinfo{volume}{166}}(\bibinfo{number}{3}), \bibinfo{pages}{791}
  (\bibinfo{year}{1968}).

\bibitem{makino20}
\bibinfo{author}{\bibfnamefont{T.}~\bibnamefont{Makino}},
  \bibinfo{author}{\bibfnamefont{K.}~\bibnamefont{Tamura}},
  \bibinfo{author}{\bibfnamefont{C.~H.} \bibnamefont{Chia}},
  \bibinfo{author}{\bibfnamefont{Y.}~\bibnamefont{Segawa}},
  \bibinfo{author}{\bibfnamefont{M.}~\bibnamefont{Kawasaki}},
  \bibinfo{author}{\bibfnamefont{A.}~\bibnamefont{Ohtomo}}, \bibnamefont{and}
  \bibinfo{author}{\bibfnamefont{H.}~\bibnamefont{Koinuma}},
  \bibinfo{journal}{Appl. Phys. Lett.}
  \textbf{\bibinfo{volume}{81}}(\bibinfo{number}{13}), \bibinfo{pages}{2355}
  (\bibinfo{year}{2002}).

\bibitem{makino24}
\bibinfo{author}{\bibfnamefont{T.}~\bibnamefont{Makino}},
  \bibinfo{author}{\bibfnamefont{K.}~\bibnamefont{Tamura}},
  \bibinfo{author}{\bibfnamefont{C.~H.} \bibnamefont{Chia}},
  \bibinfo{author}{\bibfnamefont{Y.}~\bibnamefont{Segawa}},
  \bibinfo{author}{\bibfnamefont{M.}~\bibnamefont{Kawasaki}},
  \bibinfo{author}{\bibfnamefont{A.}~\bibnamefont{Ohtomo}}, \bibnamefont{and}
  \bibinfo{author}{\bibfnamefont{H.}~\bibnamefont{Koinuma.}},
  \bibinfo{journal}{Phys. Rev. B} \textbf{\bibinfo{volume}{66}},
  \bibinfo{pages}{233305} (\bibinfo{year}{2002}).

\bibitem{kalliakos1}
\bibinfo{author}{\bibfnamefont{S.}~\bibnamefont{Kalliakos}},
  \bibinfo{author}{\bibfnamefont{X.~B.} \bibnamefont{Zhang}},
  \bibinfo{author}{\bibfnamefont{T.}~\bibnamefont{Taliercio}},
  \bibinfo{author}{\bibfnamefont{P.}~\bibnamefont{Lefebvre}}, ,
  \bibinfo{author}{\bibfnamefont{B.}~\bibnamefont{Gil}},
  \bibinfo{author}{\bibfnamefont{N.}~\bibnamefont{Grandjean}},
  \bibinfo{author}{\bibfnamefont{B.}~\bibnamefont{Damilano}}, \bibnamefont{and}
  \bibinfo{author}{\bibfnamefont{J.}~\bibnamefont{Massies}},
  \bibinfo{journal}{Appl. Phys. Lett.}
  \textbf{\bibinfo{volume}{80}}(\bibinfo{number}{3}), \bibinfo{pages}{428}
  (\bibinfo{year}{2002}).

\bibitem{hopfield1}
\bibinfo{author}{\bibfnamefont{J.~J.} \bibnamefont{Hopfield}},
  \bibinfo{journal}{Phys. Rev.} \textbf{\bibinfo{volume}{112}},
  \bibinfo{pages}{1555} (\bibinfo{year}{1958}).

\bibitem{guillaume1}
\bibinfo{author}{\bibfnamefont{C.~B.} \bibnamefont{{\`a}~la Guillaume}},
  \bibinfo{author}{\bibfnamefont{J.~M.} \bibnamefont{Debever}},
  \bibnamefont{and} \bibinfo{author}{\bibfnamefont{F.}~\bibnamefont{Salvan}},
  \bibinfo{journal}{Phys. Rev.} \textbf{\bibinfo{volume}{177}},
  \bibinfo{pages}{567} (\bibinfo{year}{1969}).

\bibitem{ohtomo8}
\bibinfo{author}{\bibfnamefont{A.}~\bibnamefont{Ohtomo}},
  \bibinfo{author}{\bibfnamefont{K.}~\bibnamefont{Tamura}},
  \bibinfo{author}{\bibfnamefont{M.}~\bibnamefont{Kawasaki}},
  \bibinfo{author}{\bibfnamefont{T.}~\bibnamefont{Makino}},
  \bibinfo{author}{\bibfnamefont{Y.}~\bibnamefont{Segawa}},
  \bibinfo{author}{\bibfnamefont{Z.~K.} \bibnamefont{Tang}},
  \bibinfo{author}{\bibfnamefont{G.}~\bibnamefont{Wong}},
  \bibinfo{author}{\bibfnamefont{Y.}~\bibnamefont{Matsumoto}},
  \bibnamefont{and} \bibinfo{author}{\bibfnamefont{H.}~\bibnamefont{Koinuma}},
  \bibinfo{journal}{Appl. Phys. Lett.}
  \textbf{\bibinfo{volume}{77}}(\bibinfo{number}{14}), \bibinfo{pages}{2204}
  (\bibinfo{year}{2000}).

\bibitem{ohtomophd}
\bibinfo{author}{\bibfnamefont{A.}~\bibnamefont{Ohtomo}},
  \emph{\bibinfo{title}{Quantum structures and ultraviolet light-emitting
  devices based on {ZnO} thin films grown by laser molecular-beam epitaxy}},
  Ph.D. thesis, \bibinfo{school}{Tokyo Institute of Technology}
  (\bibinfo{year}{2000}).

\bibitem{sun3}
\bibinfo{author}{\bibfnamefont{H.~D.} \bibnamefont{Sun}},
  \bibinfo{author}{\bibfnamefont{T.}~\bibnamefont{Makino}},
  \bibinfo{author}{\bibfnamefont{Y.}~\bibnamefont{Segawa}},
  \bibinfo{author}{\bibfnamefont{M.}~\bibnamefont{Kawasaki}},
  \bibinfo{author}{\bibfnamefont{A.}~\bibnamefont{Ohtomo}},
  \bibinfo{author}{\bibfnamefont{K.}~\bibnamefont{Tamura}}, \bibnamefont{and}
  \bibinfo{author}{\bibfnamefont{H.}~\bibnamefont{Koinuma}},
  \bibinfo{journal}{Appl. Phys. Lett.}
  \textbf{\bibinfo{volume}{78}}(\bibinfo{number}{22}), \bibinfo{pages}{3385}
  (\bibinfo{year}{2001}).

\bibitem{sunspie2001}
\bibinfo{author}{\bibfnamefont{T.}~\bibnamefont{Makino}},
  \bibinfo{author}{\bibfnamefont{H.}~\bibnamefont{Sun}},
  \bibinfo{author}{\bibfnamefont{T.~T.} \bibnamefont{Nguyen}},
  \bibinfo{author}{\bibfnamefont{Y.}~\bibnamefont{Segawa}},
  \bibinfo{author}{\bibfnamefont{C.}~\bibnamefont{Chia}},
  \bibinfo{author}{\bibfnamefont{M.}~\bibnamefont{Kawasaki}},
  \bibinfo{author}{\bibfnamefont{A.}~\bibnamefont{Ohtomo}},
  \bibinfo{author}{\bibfnamefont{K.}~\bibnamefont{Tamura}}, \bibnamefont{and}
  \bibinfo{author}{\bibfnamefont{H.}~\bibnamefont{Koinuma}}, in
  \emph{\bibinfo{booktitle}{Proceedings of the 2nd International Conference on
  Combinatorial and Composition Spread Techniques in Material and Device
  Development, San Jose}}, edited by
  \bibinfo{editor}{\bibfnamefont{G.}~\bibnamefont{Jabbour}} \bibnamefont{and}
  \bibinfo{editor}{\bibfnamefont{H.}~\bibnamefont{Koinuma}},
  \bibinfo{organization}{25 January, 2001, San Jose, USA}
  (\bibinfo{publisher}{SPIE}, \bibinfo{address}{Bellingham},
  \bibinfo{year}{2001}), vol. \bibinfo{volume}{4281}, p.~\bibinfo{pages}{68}.

\bibitem{hvam3}
\bibinfo{author}{\bibfnamefont{J.~M.} \bibnamefont{Hvam}},
  \bibinfo{journal}{Solid State Commun.}
  \textbf{\bibinfo{volume}{12}}(\bibinfo{number}{2}), \bibinfo{pages}{95}
  (\bibinfo{year}{1973}).

\end{thebibliography}
\end{document}